\documentclass[twocolumn]{aastex701}

\usepackage{amsmath}
\usepackage{wrapfig}

\newcommand{\msun}{\mathrm{M_{\odot}}}
\newcommand{\pelargir}{{\tt PELARGIR\ }}
\newcommand{\pelargirx}{{\tt PELARGIR}}
\newcommand{\pelargirs}{{\tt PELARGIR}'s\ }

\newcommand{\Sgw}{S_{\rm GW}}

\newcommand{\Nres}{N_{\rm res}}
\newcommand{\Nresj}{N_{\mathrm{res},j}}

\newcommand{\Sht}{\hat{S}_{\rm GW}}
\newcommand{\Nht}{\hat{N}_{\rm res}}

\newcommand{\thetasi}{\{\vec\theta_i\}}
\newcommand{\thetahati}{\{\hat\theta_i\}}

\newcommand{\rhothresh}{\rho_{\mathrm{thresh}}}

\mathchardef\mhyphen="2D
\graphicspath{{./}{figures/}}

\begin{document}

\title{A Foundation for Gravitational-Wave Population Inference within the LISA Global Fit}

\correspondingauthor{Alexander W. Criswell}
\author[orcid=0000-0002-9225-7756]{Alexander W. Criswell}
\affiliation{Vanderbilt University, Department of Physics and Astronomy, 2301 Vanderbilt Place, Nashville, TN 37235, USA}
\affiliation{Fisk University, Department of Life and Physical Sciences, 1000 17th Avenue N. Nashville, TN 37208, USA}
\email[show]{alexander.criswell@vanderbilt.edu}  

\author[orcid=0000-0001-7852-7484]{Sharan Banagiri}
\affiliation{Center for Interdisciplinary Exploration and Research in Astrophysics (CIERA), Northwestern University, 1800 Sherman Ave, Evanston, IL 60201, USA}
\affiliation{School of Physics and Astronomy, Monash University, VIC 3800, Australia}
\affiliation{OzGrav: The ARC Centre of Excellence for Gravitational Wave Discovery, Clayton, VIC 3800, Australia}
\email{sharan.banagiri@monash.edu}

\author[0000-0001-7099-765X]{Vera Delfavero}
\affiliation{Canadian Institute for Theoretical Astrophysics, University of Toronto, 60 St George St, Toronto, ON M5S 3H8, Canada}
\email{vera.delfavero@cita.utoronto.ca}

\author[0000-0003-0416-9818]{Maria Jose Bustamante-Rosell}
\affiliation{NASA Marshall Space Flight Center, Huntsville, AL 35812, USA}
\affiliation{Science and Technology Institute, Universities Space Research
Association, Huntsville, AL 35805, USA}
\affiliation{Vanderbilt University, Department of Physics and Astronomy, 2301 Vanderbilt Place, Nashville, TN 37235, USA}
\email{maria.jose.bustamante.rosell@nasa.gov}

\author[0000-0001-8217-1599]{Stephen R. Taylor}
\affiliation{Vanderbilt University, Department of Physics and Astronomy, 2301 Vanderbilt Place, Nashville, TN 37235, USA}
\email{stephen.r.taylor@vanderbilt.edu}

\author[0000-0002-3717-6582]{Robert Rosati}
\affiliation{NASA Marshall Space Flight Center, Huntsville, AL 35812, USA}
\affiliation{University of Alabama in Huntsville, Huntsville, AL 35805, USA}
\email{robert.j.rosati@nasa.gov}

\begin{abstract}

Population inference in gravitational-wave astronomy allows us to connect individual detections to the astrophysics of compact objects and their environments. 
Current approaches employed for population inference with LIGO-Virgo-KAGRA data approximate evaluation of the hierarchical population likelihood via post-processing of individual-event posteriors. 
However, the case of the Laser Interferometer Space Antenna (LISA) will be more complex for two main reasons: the transdimensional ``global fit'' approach to LISA data analysis which models all signals and noise simultaneously, and the presence of both individually-resolved signals and the unresolved stochastic ``Galactic foreground" arising from the Galactic binary population, which induces a circular dependence between the resolved and unresolved systems and our ability to detect the former. 
These challenges are not without opportunity; LISA's data will contain every mHz compact binary in the Milky Way --- either individually or within the Galactic foreground --- with great potential for Galactic and stellar astrophysics.
We therefore propose an alternative approach: direct evaluation of the full hierarchical population likelihood within the LISA global fit. 
We develop a statistical formalism for joint inference of individually-resolved gravitational-wave sources, an unresolved stochastic foreground, and a shared, underlying astrophysical population, present \pelargirx, a prototype GPU-accelerated population inference module for the LISA global fit,  demonstrate the formalism and \pelargir via a toy model analysis, and lay out a roadmap towards an astrophysically-motivated LISA global fit with embedded population inference. 
While we apply the formalism here to the population of LISA Galactic binaries, it is applicable across the gravitational-wave spectrum with use cases in pulsar timing and next-generation terrestrial observatories.

\end{abstract}



\section{Introduction}\label{sec:intro}

The Laser Interferometer Space Antenna (LISA) will observe a broad array of gravitational-wave (GW) sources in the mHz band. 
These include massive black hole binary mergers \citep{sesana_gravitational_2005}, extreme mass ratio inspirals \citep[EMRIs;][]{hils_gradual_1995, sigurdsson_capture_1997,amaro-seoane_intermediate_2007}, both stochastic and individual contributions from satellite dwarf galaxies of the Milky Way \citep{rieck_stochastic_2024,pozzoli_cyclostationary_2024,korol_populations_2020,keim_large_2022}, backgrounds from far-from-merger extragalactic stellar-mass binary black holes, binary neutron stars, neutron star -- black hole binaries \citep{chen_stochastic_2019, cusin_properties_2019, perigois_startrack_2021,abbott_upper_2021,lewicki_impact_2023,babak_stochastic_2023}, and white dwarf binaries \citep{staelens_likelihood_2024,boileau_gravitationalwave_2025}, individual precursors to LIGO-Virgo-KAGRA \citep[LVK;][]{aasi_advanced_2015, acernese_advanced_2015, akutsu_kagra_2019} mergers \citep{cutler_what_2019,gerosa_multiband_2019, sesana_prospects_2016, seto_how_2022}, and possibly further unknowns; see \citet{amaro-seoane_astrophysics_2023} for a review. 

Notably, LISA's data will also contain every single mHz compact binary in the Milky Way, collectively known as Galactic binaries (GBs). 
Of these tens of millions of systems, only $\sim 0.1\%$ are expected to be individually ``resolved'' sources. 
The rest will give rise to a prominent astrophysical confusion noise that will dominate LISA's instrumental noise from a few tenths of a mHz to a few mHz: ``unresolved'' GBs, or the Galactic foreground \citep{nelemans_gravitational_2001,edlund_white_2005,benacquista_consequences_2006,ruiter_lisa_2010}. 
In this work, we will take the convention that an astrophysical `foreground' refers to a stochastic astrophysical confusion noise which is greater or comparable to the instrumental or other noise sources of the observatory in question. Conversely, we will consider a stochastic GW `background' to mean the case where such a signal is subdominant to the noise.

The constant presence of an unknown number of overlapping gravitational-wave signals renders impossible independent treatment of individual signal classes and LISA's instrumental noise as is done with the current generation of ground-based detectors. 
Instead, a transdimensional ``global fit'' scheme has been proposed \citep{cornish_lisa_2005,umstatter_lisa_2005}, which seeks to model the number and nature of all source classes and the instrumental noise concurrently. Several global fit pipelines are under development \citep{littenberg_detection_2011,littenberg_global_2020, littenberg_prototype_2023,katz_efficient_2024,deng_modular_2025,strub_global_2024}, the majority of which rely on a blocked Gibbs framework \citep{geman_stochastic_1984} to perform Markov chain Monte Carlo (MCMC) sampling of the joint posterior across all GW sources and the instrumental noise.

The Gibbs schema operates in conditional blocks focused on a single model component. 
In each block, the state of all other model components (and therefore the global data residual, save for the contribution of the current source block) is held fixed. 
This allows the model to sample the \textit{conditional} posterior of the active model component, given the overall global state of all other components. 
Provided one visits each block sufficiently often and performs enough circuits around the overall framework, blocked Gibbs sampling is guaranteed to converge to the true joint posterior across all model components \citep{geman_stochastic_1984}. 
For further details on blocked Gibbs sampling as applied to the LISA global fit, see \citet{cornish_lisa_2005,littenberg_detection_2011,littenberg_global_2020, littenberg_prototype_2023,katz_efficient_2024}. 
Crucially, the number of GW sources present in the LISA data stream will not be known \textit{a priori} and must instead be inferred. 
The global fit is therefore transdimensional: both the overall number of sources and the set of possible parameters used to describe the data are allowed to vary throughout the inference process. This is accomplished via Reversible-Jump MCMC \citep[RJMCMC;][]{green_reversible_1995}.

While the necessity of this global approach to LISA data analysis is largely driven by the presence of the GBs \citep{crowder_solution_2007}, these systems --- resolved and unresolved alike --- carry significant scientific promise. 
LISA's ability to observe the Milky Way in mHz GWs has the potential to provide insights as to Galactic morphology \citep{benacquista_consequences_2006,breivik_constraining_2019}, including the as-of-yet unknown shape of the Milky Way's bar structure \citep{adams_astrophysical_2012,korol_multimessenger_2019,wilhelm_milky_2020}; metallicity \citep{yu_gravitational_2010,korol_populations_2020,thiele_applying_2021}, star formation history \citep{yu_influence_2013,korol_populations_2020}, constraints on the Milky Way initial mass function \citep{rebassa-mansergas_where_2019,korol_populations_2020}, and stellar astrophysics \citep{delfavero_recovering_2025}.

In practice, characterizing these astrophysics of interest amounts to hierarchical population inference on LISA data. Directly sampling this kind of hierarchical likelihood is computationally challenging, driven in part by the need to simultaneously infer the parameters of every member of the observed population. As such, the problem of GW population inference has traditionally been approached via a two-step procedure. 
First, one analyzes the strain data and produces a set of posterior samples for each observed event. 
Then, one can use importance sampling or other techniques to infer the characteristics of the underlying population, while accounting for the selection effects imposed by the incomplete observational sample \citep{loredo_accounting_2004, mandel_extracting_2019, fishbach_does_2018, Essick:2023upv}. 
This is the procedure used for population inference on LVK stellar-mass compact binary mergers \citep[e.g.,][]{LIGOScientific:2025jau,Galaudage:2024meo, Adamcewicz:2025phm, Banagiri:2025dxo, Antonini:2025zzw, Tong:2025wpz, Callister:2022qwb,McFACTSIII, Ford_2026,LIGOScientific:2025pvj, Banagiri:2025dmy,  Plunkett:2026pxt, Wang:2025nhf, Colloms:2025hib, Galaudage:2024meo}. 
A similar two-step procedure has also been used for inference of the supermassive black hole binary population from the stochastic nHz GW background \citep{harris_connecting_2024,chen_inference_2026,bi_implications_2023, ellis_gravitational_2024, liepold_big_2024, sato-polito_where_2024,agazie_nanograv_2023c}.\footnote{It is worth noting that by our definition, the nHz GW background in fact qualifies as an astrophysical foreground.} 
Simultaneous hierarchical inference of individual and stochastic signals of common astrophysical origins was developed in \citet{callister_shouts_2020}, which extended this two-step approach to the case of joint inference of resolved LVK mergers and an astrophysical GW background arising from the same population \citep[see also][which builds on that work]{cousins_stochastic_2026}. 

There are two key differences between the LVK case and the LISA GB case. 
First is the matter of selection effects (or lack thereof). 
In the LISA case, the population considered is (frequency) complete. 
All members of the population are included in some aspect of the analysis, either individually as resolved sources or collectively via their contributions to the unresolved foreground. 
As a result, no selection effects are present provided the population is treated holistically \citep[see, e.g., discussion in][]{thrane_measuring_2013,smith_optimal_2018}.
The second is our inability with LISA to cleanly separate the data containing resolvable signals as needed in the joint inference framework of \citet{callister_shouts_2020}. 
For LISA, our ability to resolve individual systems is dependent on the level of the Galactic foreground power spectral density (PSD). 
Conversely, because the foreground is definitionally composed of those sources which we cannot individually resolve, the foreground PSD is dependent on our ability to resolve individual sources and remove them from the unresolved categorization. This is a fundamentally circular logic. The hierarchical population model is therefore circularly-dependent; the resolved and unresolved components of the likelihood cannot be treated as separable. 
As such, one must jointly infer the resolved systems, unresolved foreground, and the underlying population in a holistic fashion.

 Recently, \citet{toubiana_framework_2026} proposed a means to approach this problem via a variation of the traditional two-step method. 
By invoking a user-defined ``resolvability function'' and treating the resolved GBs and the unresolved foreground as arising from a single, underlying population, they are able to infer population hyperparameters by applying importance sampling techniques to a toy model global fit posterior. However, as the resolvability function is not astrophysically-motivated, this approach requires an external choice for its form and parameters. As \citet{toubiana_framework_2026} note, this can result in complications for the population inference scheme if the choice of resolvability function is ill-suited to the data.
Moreover, this kind of in-post approach may struggle in the context of the full $\mathcal{O}(100,000)$-parameter, transdimensional global fit. 
Techniques based on importance sampling and Monte Carlo approximations to hierarchical inference can encounter challenges related to convergence and reweighting efficiency. 
While several studies have focused on developing sophisticated techniques to estimate and handle such issues in the LVK context of $\mathcal{O}(100\mathrm{s})$ of signals \citep{Farr_2019, Talbot:2023pex, Essick:2022ojx, Heinzel:2025ogf}, they have also found that the induced error scales with the number of events considered \citep{Farr_2019, Talbot:2023pex}.

Additionally, the transdimensional nature of the LISA global fit can be expected to greatly exacerbate any issue of reweighting efficiency. 
This is in part indicated by the prior-dependence of the inferred resolved GBs in the case of multiple global fits applied to a common dataset. 
When applied to LISA Data Challenge: Sangria \citep{lejeune_lisa_2022}, each global fit arrived at statistically different solutions \citep[cf.][]{katz_efficient_2024,littenberg_prototype_2023,deng_modular_2025}.\footnote{This may also be an issue of convergence; it is difficult to show that current global fit prototypes are fully converging. If this is the case, the effects discussed here will be further exacerbated due to lack of posterior coverage in low-probability regions.} \citet{littenberg_prototype_2023} noted a severe impact on source resolvability when moving from a (static) galaxy-shaped prior on the GB spatial distribution to a uniform one. 
This kind of prior-dependence may result in near-zero or zero reweighting efficiency for traditional, in-post approaches to population inference with LISA, even \citep[and perhaps especially, given the results of][]{littenberg_prototype_2023} in the case of reweighting from a broad, ``agnostic'' prior to a population-informed one.

Finally, \citet{srinivasan_simulationbased_2025} and \citet{santi_inferring_2026} recently developed frameworks for neural posterior estimation trained on simulated LISA data. \citet{santi_inferring_2026} consider only the Galactic foreground --- removing resolved sources from their training data via an iterative subtraction algorithm --- and show that the foreground spectrum alone carries useful information as to the population. In contrast, \citet{srinivasan_simulationbased_2025} avoid the issue of circularity entirely, approximating (and therefore inverting) the connection between the population parameters and a full simulated LISA datastream via a normalizing flow architecture.

In this work, we instead develop the approach of directly sampling of the hierarchical population likelihood by embedding GB population inference within the LISA global fit itself. 
While a computationally-challenging endeavor, the major hurdle --- simultaneously inferring the total number and parameters of some 10,000 resolved GBs alongside the Galactic foreground and the LISA instrumental noise at the level of the data --- is already within reach of current global fit prototypes. 
As such, incorporating GB population inference within the global fit amounts to the inclusion of a single additional block within the overall blocked Gibbs framework. The global residual would remain unchanged within such a block. Instead, a population inference block updates the overall noise covariance (via the foreground PSD) and the priors of the resolved GB block.
Crucially, such an approach inherently bypasses all issues of reweighting accuracy and efficiency. 

This is not the first effort to sample a full hierarchical likelihood in the context GW population inference. 
\citet{mancarella_sampling_2025} considered the LVK case and simultaneously sampled the population model's hyperparameters alongside the individual event posteriors. 
However, they still relied on a two-step approach based on the individual-event posteriors rather than the raw LIGO strain data, avoiding importance sampling via Gaussian Mixture Model emulation of the single-event posteriors. 
\citet{laal_deep_2025} evaluated the full hierarchical likelihood at the level of the data in order to simultaneously infer characteristics of the supermassive black hole binary population and the nHz GW background via neural emulation of a semi-analytic population model.  
To date, however, no similar method has been put forth for the case considered in this work of resolved systems in the presence of an unresolved foreground. To this end, we present a general statistical framework, suitable for \textit{in situ} application within a global fit, to perform hierarchical population inference of resolved GW signals alongside an astrophysical foreground arising from the same source population. 

We then take the first step towards a treatment of the problem directly within the LISA global fit. 
This approach relies on rapid sorting --- within each likelihood evaluation --- of millions of GBs drawn from a population model into resolved and unresolved systems, thereby allowing for direct calculation of the Poisson statistics of the resolved binaries as well as the resulting PSD of the Galactic foreground.
This framework is implemented within \pelargirx, a GPU-accelerated prototype module for GB population inference within the LISA global fit. We demonstrate the formalism and \pelargir via a simplified toy model, lay out a roadmap for inference of population astrophysics within the LISA global fit, and discuss applications of this framework across the spectrum of GW astronomy.

\section{A Formalism for Gravitational-Wave Population Inference in the Presence of Astrophysical Foregrounds }\label{sec:formalism}

Here we present a formalism for joint, hierarchical inference of a set of resolved systems, an unresolved astrophysical foreground, and the characteristics of their shared underlying population. 
While we will in this work primarily consider the population of Galactic white dwarf binaries, the formalism presented here can be extended to any relevant population requiring joint inference of its resolved and unresolved components, including the supermassive black hole binary population in pulsar timing arrays. 

Suppose a population of GBs --- resolved and unresolved alike --- such that the Milky Way contains $N$ compact mHz binaries within the LISA frequency band. Each GB is characterized by some set of astrophysical parameters $\vec\theta$. We wish to infer the characteristics of the overall population, given by population hyperparameters $\Lambda$, such that $\Lambda$ describes the joint prior on $\vec\theta$.
From this population, let some $\Nres = \Nres(N,\Lambda)$ of these signals be resolved, with the $i^{\rm th}$ resolved GB being described by parameters $\vec\theta_i$ for $i \in 1...\Nres$. The remaining $N-\Nres$ GBs comprise the unresolved astrophysical foreground. This is an anisotropic stochastic GW signal whose PSD and anisotropy are both determined by the population such that its anisotropic PSD as a function of frequency and sky direction is
\begin{equation}
    \Sgw = \Sgw(f,\hat\Omega|N,\Lambda)\,.
\end{equation}
Note that $\Nres$ and $\Sgw$ are circularly-dependent as discussed in \S\ref{sec:intro}. $\Nres$ is a point estimate at this step, but becomes a distribution through uncertainties in the hyperparameters.
We write the LISA instrumental noise covariance for a given set of time-delay interferometry \citep[TDI;][]{tinto_timedelay_2020,tinto_secondgeneration_2023} channels --- either X-Y-Z or A-E-T --- as $C_n(f;t|\eta)$ for an arbitrary noise model with parameters $\eta$. 
This allows us to construct the overall covariance $C$ of LISA noise, combining the astrophysical and instrumental contributions,
\begin{equation}\label{eq:Cdef}
    C(f;t|N,\Lambda,\eta) = C_n(f;t|\eta) + C_{\rm GW}(f;t|N,\Lambda)\,,
\end{equation}
where $C_{\rm GW}(f;t|N,\Lambda)$ is the induced covariance across TDI channels from the foreground such that
\begin{equation}
    C_{\rm GW}(f;t|N,\Lambda) = \int\mathcal{R}(f,\hat{\Omega};t)\Sgw(f,\hat{\Omega}|N,\Lambda)d\hat{\Omega}\,,
\end{equation}
with $\mathcal{R}(f,\hat{\Omega};t)$ denoting the time- and frequency-dependent directional LISA response functions to incident GWs (see, e.g., \citet{romano_detection_2017} for a review). Note that $C_n(f;t|\eta)$, $C_{\rm GW}(f;t|N,\Lambda)$ and therefore $C(f;t|N,\Lambda,\eta)$ are all $3 \times 3$ covariance matrices.

We can now proceed to the overall hierarchical formalism. 
We wish to infer the joint posterior distribution of the Galactic foreground PSD including its anisotropy $\Sgw(f,\hat{\Omega})$, the resolved GB parameters $\thetasi$, the number of resolved binaries $\Nres$,\footnote{In practice, it will be more useful to parameterize and infer the number of resolved GBs in each frequency bin $\Nresj(f_j)$ for $j\in1...N_f$. Extending this formalism for this use case is trivial as $\Nres = \sum_j\Nresj$ and the sum of Poisson variables is itself Poisson-distributed.} the instrumental noise model parameters $\eta$ and the population hyperparameters $\Lambda$ and $N$. 
We therefore write the following via Bayes' theorem for some data $d$:
\begin{widetext}
    \begin{equation}\label{eq:initial_bayes}
    p(\Lambda,N,\Sgw(f,\hat{\Omega}),\Nres,\thetasi,\eta|d) = \frac{\mathcal{L}(d|\Lambda,N,\Sgw(f,\hat{\Omega}),\Nres,\thetasi,\eta)}{p(d)}\pi(\Lambda,N,\Sgw(f,\hat{\Omega}),\Nres,\thetasi,\eta)\,.
\end{equation}
Under the assumption of Gaussianity (though this need not necessarily be the case), the likelihood in Eq.~\eqref{eq:initial_bayes} is an extension of the Whittle likelihood \citep{whittle1951hypothesis} to the complex multivariate Gaussian case \citep{adams_discriminating_2010}. It only depends on the population through the observables:
    \begin{equation}\label{eq:whittle_likelihood}
        \mathcal{L}(d|\Sgw,\Nres,\thetasi,\eta)= \frac{1}{2\pi T_{\rm seg}|C|}\exp{\left\{ -\frac{2}{T_{\rm seg}} \left[\tilde{d}(f) - \sum_i^{N_{\rm res}} \tilde{h}_i (f | \vec\theta_i)\right]^{\dagger} C^{-1} \left[\tilde{d}(f) - \sum_i^{N_{\rm res}} \tilde{h}_i (f | \vec\theta_i)\right] \right\}}\,,
\end{equation}
\end{widetext}
where $C=C(f;t|N,\Lambda,\eta)$ is defined as per Eq.~\eqref{eq:Cdef} and includes the foreground contribution, and $\tilde{h}_i(f|\theta_i)$ is the Fourier-domain strain of an individual GB. 
This form assumes a short-time Fourier transform approach with time-segment length $T_{\rm seg}$; however, other choices such as a time-frequency representation \citep[e.g.,][]{cornish_nonstationary_2025} could be made without loss of generality. 
The joint prior in Eq.~\eqref{eq:initial_bayes} can be factored as follows:
\begin{align}
        \pi(& \Lambda,N, \Sgw,\Nres,\thetasi,\eta) \nonumber \\
        &\qquad= \pi(\Sgw,\Nres,\thetasi|N,\Lambda)\pi(N,\Lambda)\pi(\eta)\,,
\end{align}
where we have dropped the arguments of $\Sgw$ for simplicity.
Neglecting the evidence $p(d)$ and asserting that the only dependence of the likelihood on the population is via $\Nres$, $\Sgw$, and $\thetasi$, the full expression then becomes
\begin{align}\label{eq:full_posterior}
            p(&\Lambda,N,\Sgw,\Nres,\thetasi,\eta|d) \nonumber\\
            &\propto \mathcal{L}(d|\Sgw,\Nres,\thetasi,\eta)\nonumber\\
            &\quad\times\pi(\Sgw,\Nres,\thetasi|N,\Lambda,\eta)\pi(\eta)\pi(N,\Lambda)\,,
\end{align}
which if desired can be marginalized over the resolved binary parameters, the total number of resolved binaries, the unresolved stochastic foreground, and the instrumental noise to yield the posterior distribution of our population parameters alone:
\begin{widetext}
    \begin{equation}\label{eq:full_marginal_posterior}
\boxed{
    p(\Lambda,N|d) \propto \iiiint\mathcal{L}(d|\Sgw,\Nres,\thetasi,\eta)\pi(\Sgw,\Nres,\thetasi|N,\Lambda,\eta)\pi(\eta)\pi(N,\Lambda)d\Sgw d\Nres d\thetasi d\eta\,.
}
\end{equation}
\end{widetext}
In contrast to \citet{abbott_binary_2016,fishbach_does_2018, fishbach_where_2017,mandel_extracting_2019}, there is no explicit term in Eq.~\eqref{eq:full_marginal_posterior} governing the detection probability of a system with specific parameters $\vec\theta$. 
This is because we consider a complete sample of the population; all systems fall into either the resolved or unresolved category and are therefore considered in one fashion or another.
Selection effects arise from discarded data/events, of which we have none. The term $\pi(\Sgw,\Nres,\thetasi|N,\Lambda,\eta)$ therefore fully encapsulates the population model. It must necessarily contain the circular dependence of the resolved and unresolved systems discussed in \S\ref{sec:intro} and the associated joint dependence of $\Sgw$, $\Nres$, and $\thetasi$. 

\subsection{Semi-Analytic Population Model}\label{sec:formalism_sams}
Here we develop a semi-analytic model for estimation of $\pi(\Sgw,\Nres,\thetasi|N,\Lambda,\eta)$, including its intrinsic circularity. In order to calculate the mapping between the population hyperparameters and the observables, we treat the problem at the level of the population's constituent binaries. We draw a full Galaxy of N binaries per realization from the population-level priors given by $\Lambda$ and apply a thresholding procedure (described below) to divide them in a self-consistent fashion between resolved and unresolved systems. As the semi-analytic model provides estimators for the observable quantities $\Sgw$, $\Nres$, and $\thetasi$, let $\Sht$, $\Nht$, and $\thetahati$ be the result of a semi-analytic model evaluation

Given a population described by $\Lambda$, both observed reality and the output of the semi-analytic model are different Poisson\footnote{This is a spatial Poisson point process with a volumetric rate, rather than the more familiar temporal case.} realizations of the same underlying process. Due to the circularity of the problem, the intrinsic rate of these Poisson processes cannot be analytically derived from the population priors. As such, the correct approach is to calculate the probability of the observed realization given our model realizations, marginalized over the shared underlying process. That is, we should expand $\pi(\Sgw,\Nres,\thetasi|N,\Lambda,\eta)$ as
\begin{widetext}
    \begin{equation}
    \pi(\Sgw,\Nres,\thetasi|N,\Lambda,\eta) = \iint p(\thetasi|N,\Lambda,\eta) p(\Sgw|\Sht)p(\Nres|\Nht)p(\Sht,\Nht|N,\Lambda,\eta)d\Nht d\Sht\,,
    \label{eq:population-joint}
\end{equation}
\end{widetext}
where $p(\Sht,\Nht|N,\Lambda,\eta)$ is described by our semi-analytic model, and $p(\Sgw|\Sht)$ and $p(\Nres|\Nht)$ are the necessary marginal terms discussed above. 

We now lay out a thresholding procedure to estimate $p(\Sht,\Nht|N,\Lambda,\eta)$. 
We derive the result for a single frequency bin, which can then be generalized across all frequency bins. 
Let $\rhothresh$ be some SNR threshold which divides the resolved and unresolved binaries in frequency bin $j$. 
This quantity need not be the same in every frequency bin, nor need it be fixed; it can --- and, in practice, should --- be allowed to vary and be inferred as a model parameter or marginalized over. 
We can then write formal expressions for $\Nht(f_j)$ and $\Sht(f_j)$ -- the number of resolved binaries and the foreground PSD in frequency bin $j$, respectively -- given a draw of $N$ GBs from $p(\vec\theta|\Lambda)$:
\begin{widetext}
\begin{align}\label{eq:sam_nres_int}
    \Nht(f_j;N,\Lambda,\eta,\rhothresh) & = N\int p(\vec\theta|\Lambda) \Theta\left\{\rho(\vec\theta_i)  \geq \rho_{\rm thresh} \right\}\delta(f(\vec\theta)-f_j)d\vec\theta\\
    \label{eq:sam_nres_discrete}
    & = \sum_{i=1}^N \Theta\left\{\rho(\vec\theta_i)  \geq \rho_{\rm thresh} \right\}\delta(f_i-f_j)\,,\\
    \label{eq:sam_sgw_int}
    \Sht(f_j;N,\Lambda,\eta,\rhothresh) & = N\int p(\vec\theta|\Lambda)S(\vec\theta)\Theta\left\{\rho(\vec\theta)  < \rho_{\rm thresh} \right\}\delta(f(\vec\theta)-f_j)d\vec\theta\\
    \label{eq:sam_sgw_discrete}
    & = \sum_{i=1}^N S(\vec\theta_i)\Theta\left\{\rho(\vec\theta_i) < \rho_{\rm thresh} \right\}\delta(f_i-f_j)\,,
\end{align}
\end{widetext}
wherein $f$ and $f_i$ are functions of $\{m_1, m_2, a\}$, $\Theta$ is the Heavyside step function and $S(\vec\theta)$ is the PSD contribution of an individual binary. 
The SNR $\rho$ determines, for a given binary, if it is treated as resolved or unresolved within the formalism. It is defined in terms of the astrophysical parameters of that binary, the population hyperparameters, and the parameters of the instrumental noise model:
\begin{widetext}
    \begin{equation}\label{eq:snr}
    \rho^2(\vec\theta;f,t,C)= 4 \iint \, \tilde{h}^{\dagger} (f,t|\vec\theta) C^{-1}(f, t | N,\Lambda, \eta) \tilde{h}(f,t|\vec\theta)dfdt\,,
\end{equation}
\end{widetext}
where the covariance matrix is given by Eq.~\eqref{eq:Cdef}.
In a practical setting with finite frequency resolution, the delta functions should be taken to mean
$$f_j-\frac{\Delta f}{2}< f_i \leq f_j+\frac{\Delta f}{2}\,,$$
where $\Delta f$ is the frequency resolution, which is to say that $f_i$ falls in frequency bin $j$. 

Note the circularity present in Eqs.~\eqref{eq:sam_nres_int}--\eqref{eq:snr}: the SNR $\rho$ is a function of both the individual binary parameters and the covariance $C$, which is in turn a function of $\Sht$, which is dependent on the individual binary parameters via the division of resolved and unresolved binaries. We develop a quick non-iterative estimator for these quantities as follows. 
First, we calculate the ``na\"ive'' SNR $\rho_n$, i.e., the SNR with respect to only the instrumental noise,
\begin{equation}
    \rho_{\mathrm{n}}(\vec\theta;f,t) \equiv \rho(\vec\theta;f,t,C=C_n)\,.
\end{equation}
Then, all $N_j$ GBs within a given frequency bin $j$ are ordered from lowest to highest na\"{i}ve SNR. This allows us to calculate the cumulative SNR estimator $\hat\rho_c$ for each GB such that  $\hat\rho_{c,i} =\rho(\vec\theta_i;f_j,t,\hat{C}_{ij})$,  where the cumulative noise covariance estimator for binary $i$ in frequency bin $j$, $\hat{C}_{ij}$ is:
\begin{equation}
    \hat{C}_{ij} = C_n(f_j|\eta) + \sum_{k=0}^{i-1}\mathcal{R}(f_j,\hat{\Omega}_k;t)S(\vec\theta_k)\,.
    \medskip
\end{equation}
That is, the cumulative SNR estimator $\hat\rho_{c,i}$ gives the nominal SNR of binary $i$ with respect to both the instrumental noise and the astrophysical noise contribution of every system with a lower na\"ive SNR. 
Note that  $\hat\rho_c$ as computed in this fashion is not necessarily monotonic; there can exist systems whose SNR estimator nominally exceeds $\rhothresh$ when only considering lower na\"ive SNR systems, but for which there exists at least one system with a greater na\"ive SNR which is itself subthreshold. 
If not all systems above the system in question can be resolved, it will in turn not be resolved. 
As such, the true boundary in SNR between the resolved and unresolved systems occurs at the system with the largest na\"ive SNR which cannot be resolved from the systems below it:
\begin{equation}
    \hat\rho_{\mathrm{boundary}} \equiv \max \left(\rho_{n,i}\ |\ \hat\rho_{c,i} < \rhothresh \right)\,.
\end{equation}
This leads to the final, practical expressions for $\Nht$ and $\Sht$ in terms of the SNR estimator $\hat\rho_c$:
\begin{widetext}
\begin{equation}\label{eq:sam_nres_practical}
    \Nht(f_j;N,\Lambda,\eta,\rhothresh) = \sum_{i=1}^N \Theta\left\{\hat\rho(\vec\theta_i)  > \hat\rho_{\mathrm{boundary}} \right\}\delta(f_i-f_j)\,,
\end{equation}
and
\begin{equation}\label{eq:sam_sgw_practical}
    \Sht(f_j;N,\Lambda,\eta,\rhothresh) = \sum_{i=1}^N S(\vec\theta_i)\Theta\left\{\hat\rho(\vec\theta_i) \leq \hat\rho_{\mathrm{boundary}} \right\}\delta(f_i-f_j)\,.
\end{equation}
\end{widetext}

We can now connect the result of the semi-analytic model evaluations to the observables, via the marginal terms $p(\Sgw|\Sht)$ and $p(\Nres|\Nht)$. In principle, these can also be sampled over and estimated via a Monte Carlo integral. Ideally, however, we would like to compute these terms analytically so as to make their evaluation efficient in practice.

In appendices \ref{appendix:formalism_marginal_poisson} and~\ref{appendix:formalism_marginal_gauss}, we derive a conjugate-prior-based approach which allows for analytic marginalization over the unknown underlying process while accounting for the uncertainty induced by doing so via a finite set of realizations from the semi-analytic model. 
To do so, we define $\{\Nht\}_r$ and $\{\Sht\}_r$ to be a set of $N_r$ evaluations of the semi-analytic thresholding procedure as described above, such that the analytic marginal terms given multiple realizations are written as $p(\Nres|\{\Nht\}_r)$ and $p(\Sgw|\{\Sht\}_r)$.
We consider two cases, one for pure Poisson statistics, and the second for the Gaussian limit of many sources. The result for each case is as follows: $p(\Nres|\{\Nht\}_r)$ is described by a negative binomial as given in Eq.~\eqref{eq:marginal_negative_binomial}, and $p(\Sgw|\{\Sht\}_r)$ is described by a location-scale Student's t distribution as defined in Eq.~\eqref{eq:marginal_t}. 

With all these terms in hand, we can estimate the final remaining term, $p(\thetasi|N,\Lambda,\eta)$, by considering the average detection probability conditioned on the semi-analytic model realizations:
\begin{widetext}
    \begin{equation}\label{eq:sam_pthetas_practical}
\begin{split}
    p(\thetasi|N,\Lambda,\eta) &= \sum_{i=1}^{\Nres} \left[\pi(\vec\theta_i|\Lambda)\,p(\mathrm{resolved}|\vec\theta_i,N,\Lambda,\eta)\right]\\
    &\simeq \sum_{i=1}^{\Nres} \left[\pi(\vec\theta_i|\Lambda)\,\frac{1}{N_r}\sum_{k=1}^{N_r} \Theta\left\{ \hat\rho\left(\vec\theta_i,\eta,\{\Sht\}_r\right) \geq \rho_{\rm thresh} \right\}\right]\,.
\end{split}
\end{equation}
\end{widetext}

We now have all the pieces for the complete posterior for joint inference of resolved and unresolved systems under a semi-analytic population model as given by Eq.~\eqref{eq:full_marginal_posterior}.
\begin{itemize}
    \item The GW likelihood is given by Eq.~\eqref{eq:whittle_likelihood}.
    \item The semi-analytic estimators $\Nht$ and $\Sht$ under the population parameters $\{N,\Lambda\}$ are given by Eqs.~\eqref{eq:sam_nres_practical} and~\eqref{eq:sam_sgw_practical}, respectively.
    \item The marginal probability of $\Nres$ and $\Sgw$ given a set of semi-analytic model realizations are given by Eqs.~\eqref{eq:marginal_negative_binomial} and~\eqref{eq:marginal_t}, respectively.
    \item The probability of the observed resolved GB parameters $\thetasi$, conditioned on the semi-analytic estimators, is given by Eq.~\eqref{eq:sam_pthetas_practical}.
\end{itemize}

This procedure allows for astrophysically-motivated joint estimation of a set of resolved systems and their counterpart unresolved astrophysical foreground given a threshold SNR. From a practical, computational perspective, the calculations described are straightforward to implement as parallelizable array operations well-suited to GPU-accelerated evaluation. The formalism can also be extended for a heterogeneous population consisting of several interconnected subpopulations as will in fact be the case for LISA; we present a derivation for this case in Appendix~\ref{appendix:multipop}.

\section{PELARGIR}\label{sec:pelargir}


We implement the formalism described in this work as a foundation for the first prototype population inference module for the LISA global fit, dubbed \pelargir (Population Estimation for LISA in A Reversible-jump Global Inference Regime). 
The code is CPU/GPU agnostic and written in Numpy \citep{harris_array_2020} and Cupy \citep{okuta_cupy_2017}. 
\pelargir has been designed to be integrated within a global fit environment, with flexible class structures to support iterative improvements in astrophysical modelling and global fit integration and native support for the transdimensional sampler {\tt Eryn} \citep{karnesis_eryn_2023}, which is used in the GPU-accelerated Erebor global fit \citep{katz_efficient_2024}. 
\pelargir is open source and can be found on GitHub: \href{https://github.com/criswellalexander/pelargir-gb}{https://github.com/criswellalexander/pelargir-gb}. 

\subsection{Thresholding Module}\label{sec:pel_thresher}

A major computational challenge in order to implement the proposed framework in a Global Fit setting is rapid handling of the procedure outlined in \S\ref{sec:formalism_sams}. Prior to this work, there existed two main means by which one could compute the resolvability of GBs from the Galactic foreground:
the iterative subtraction algorithm of \citet{karnesis_characterization_2021} and, of course, running a Global Fit. The latter is infeasible to do recursively (i.e., running a global fit at every iteration of the global fit). The former approach consists of iteratively removing GBs with SNR above some threshold, computing the new foreground level, and repeating until no resolvable binaries remain. The original implementation had a typical runtime of $\mathcal{O}$(hrs), and while a recent GPU-accelerated update \citep{santi_inferring_2026} has reduced this to $\mathcal{O}$(min), neither approach is sufficiently rapid so as to be suitable for repeated execution in an inference setting. 

\begin{figure*}
    \centering
    \includegraphics[width=0.8\linewidth]{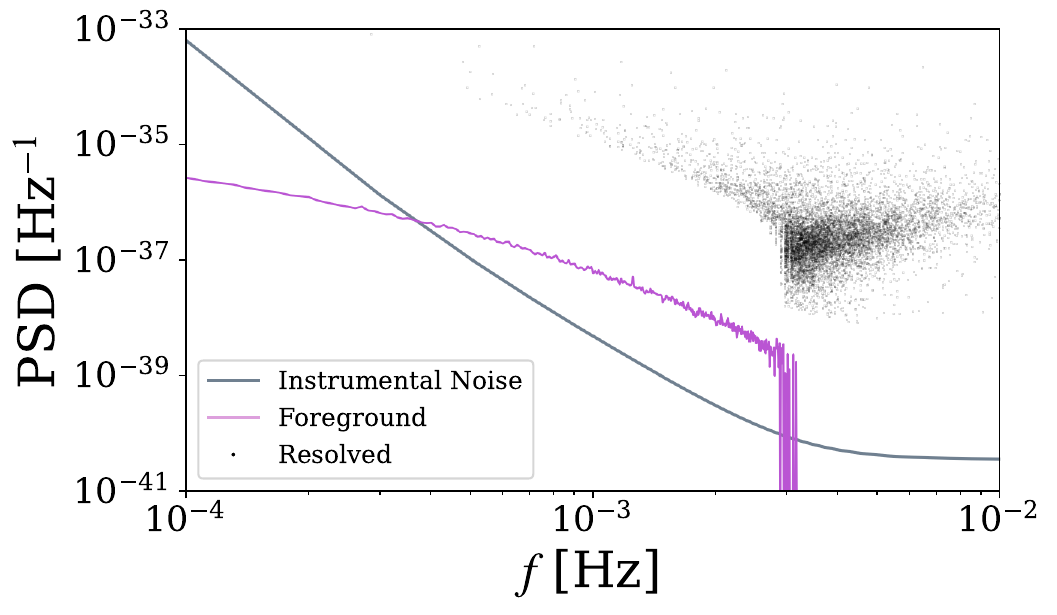}
    \caption{Result of the thresholding algorithm as applied to the fiducial population synthesis catalogue of \citet{thiele_applying_2023} at a frequency resolution of $\Delta f = 10^{-5}$ Hz, representative of the resolution of short-time Fourier transform based approaches. \pelargir finds $\Nres=8,091$ resolved systems, depicted in black. }
    \label{fig:sorting-algo-demo}
\end{figure*}

To this end, we implement the prescription described in \S\ref{sec:formalism_sams} as a rapid array sorting module, {\tt PELARGIR.thresholding}. 
The module is parallelized across frequencies, realizations, and parallel likelihood evaluations. 
The result of this module as applied to the fiducial population synthesis catalogue of \citet{thiele_applying_2023} is shown in Fig.~\ref{fig:sorting-algo-demo} and takes $6.0\pm0.1$ s to evaluate at a frequency resolution of $\Delta f =10^{-5}$ Hz. While this exceeds the speed of other methods such as iterative subtraction, further optimization is needed, likely via machine learning emulation. For the frequency resolution and limits considered in the toy model, the Thiele catalogue has an evaluation time of $1.4\pm0.3$ s. This module has applications well beyond population inference alone, and can serve as a fast and straightforward tool for estimating the resolved and unresolved components of simulated GB populations.

\section{Toy Model}\label{sec:toymodel}

We now validate our formalism for population inference via a simplified toy model of a single global fit block evaluation. 
For this demonstration, we make a series of simplifying assumptions.
\begin{itemize}
    \item \pelargir receives a single, fixed state of the resolved GB parameters, equivalent to demonstrating the model as a block within the Global Fit receiving the current state of the GB sampler. This additionally means that the ``observed'' $\Nres$ and $\thetasi$ remain fixed throughout the analysis.
    \item Consequently, we cannot properly marginalize over the uncertainty on the resolved GB parameters, as would naturally occur within the global fit. We therefore fix the input GB parameters to their simulated values to avoid inducing bias from neglecting this uncertainty.
    \item We fix $\rhothresh=7$, although we emphasize that making this kind of choice is not necessary within the formalism.
    \item We do not treat anisotropy, instead assuming a flattened 1D distance distribution.
    Astrophysically-driven anisotropic modelling of the Galactic foreground is not trivial \citep[see, e.g., discussion in][]{criswell_templated_2025}.
    \item For efficiency in testing, we set $f_{\rm min}=0.3$ mHz, allowing us to simulate a realistic foreground with only $N=10^6$ GBs (as the vast majority of GBs by number lie at low frequencies).
    \item We neglect any detailed signal processing and take the assumption of direct access to an abstracted log-Normal likelihood on the overall residual PSD (including both the instrumental noise and astrophysical foreground contributions). 
\end{itemize}

\subsection{Population Model}
We model the GB population in terms of its mass distribution, orbital separation distribution, and a flattened 1D distance distribution. Each binary is therefore modelled in terms of $\vec\theta \equiv \{m_1,m_2,a,d_L\}$, i.e. the individual binary masses, binary orbital separation, and luminosity distance. 
We make the simplifying assumption that all binaries are circular, non-evolving, and therefore monochromatic. 
We take the individual binary masses  to be independent and Normally-distributed, such that for ${\rm X} \in \{1,2\}$
\begin{equation}
    p(m_{\rm X}) \sim \mathcal{N}(\mu_m,\sigma_m)\,,
\end{equation}
where $\mu_m$ and $\sigma_m$ are the mass mean and variance. This distribution is truncated at $m_{\rm min}=0.17\,\msun$ \citep[the lowest-observed white dwarf mass in the survey of][]{kilic_lowest_2007}\footnote{For the sake of completeness, we note the existence of the $0.07\,\msun$ white dwarf observed in an AM Canum Venaticorum system \citep{boneva_midcycle_2020}, but do not otherwise take it into account.} and $m_{\rm max}=1.44\,\msun$ \citep[the Chandrasekhar limit;][]{chandrasekhar_maximum_1931}. The orbital separation distribution is modelled as a power law with spectral index $\alpha_a$, such that
\begin{equation}
    p(a) \sim a^{\alpha_a}\,,
\end{equation}
with a minimum allowed orbital separation at $10^{-4}$\,AU and a maximum at $10^{-2}$\,AU (to avoid simulating systems outside the LISA band).
We construct a simplified 1D distance distribution, consisting of a Gaussian bulge and an exponential disk, flattened into 1D and truncated to not allow unreasonably nearby binaries. This is parameterized as
\begin{equation}
\begin{split}
    p(d_{\rm L}) &\sim q_{\rm BD}\times\mathcal{N}(8\, \mathrm{kpc},r_{\rm bulge}) \\
    &\quad + (1-q_{\rm BD})\times\mathrm{Exponential}(1/r_{\rm disk})\,,
\end{split}
\end{equation}
where $r_{\rm bulge}$ is the characteristic radius of the bulge, $r_{\rm disk}$ is the disk radial scale height, and $q_{\rm BD} = M_{*,\rm bulge}/M_{*,\rm disk}$ is the bulge-to-disk stellar mass ratio. 
The exponential disk is symmetric about 8\,kpc and wraps in $d_L$ at the position of Earth. No system is allowed to be closer than the nearest star ($d_{\rm L,min}=1$\,pc).
The set of population hyperparameters $\Lambda$ is then $\Lambda \equiv \{\mu_m,\sigma_m,\alpha_a,r_{\rm bulge},r_{\rm disk},q_{\rm BD}\}$. The corresponding hyperpriors are given in Table~\ref{tab:hyperpriors_truevals}.

\begin{table*}
    \centering
    \begin{tabular}{|c|c|c|c|}
    \hline
         Hyperparameter & Hyperprior & Simulation Value & [Unit]\\
         \hline\hline
         $\mu_m$ & $\mathcal{U}(0.2,1.1)$ & 0.6 & $M_{\odot}$\\
         $\sigma_m$ & $\mathrm{InvGamma}(a=7)$ & 0.15 & $M_{\odot}$\\
         $\alpha_a$ & $\mathcal{U}(-0.5,1.5)$ & 0.5 & $-$\\
         $r_{\rm disk}$ & $\mathcal{U}(1,10)$ & 3.31 & kpc\\
         $r_{\rm bulge}$ & $\mathcal{U}(0.05,2)$ & 0.75 & kpc\\
         $q_{\rm BD}$ & $\mathcal{U}(0.01,0.99)$ & 0.33 & $-$\\
     \hline
    \end{tabular}
    \caption{Hyperpriors and simulated values for the simple population model.}
    \label{tab:hyperpriors_truevals}
\end{table*}

\subsection{Toy Model Analysis}
We simulate a toy model dataset with true (simulated) population parameters as follows. 
The mass distribution parameters are set to $\mu_m=0.6\,\msun$ and $\sigma_m=0.15\,\msun$ to mimic the observed distribution of field white-dwarf masses \citep[see e.g.,][]{tremblay_field_2016}. 
The orbital separation power law slope is set to $a_{\alpha}=0.5$ such that the corresponding low-frequency slope of the Galactic foreground spectrum will be $7/3$ in $\Sgw$, as expected for a stochastic signal arising from a binary population under purely gravitaitonal-wave driven evolution prior to the impacts of source discreteness \citep{phinney_practical_2001,sesana_stochastic_2008}. 
Taking the Galactic bulge to be Gaussian in 1D is a simplification, so setting the characteristic radius to be $r_{\rm bulge}=0.75$ kpc is a somewhat arbitrary choice. Nonetheless, this results in the majority of the bulge contribution sitting within a $\sim2-3$\,kpc radius of the Galactic center, as expected. 
The bulge is known to contain $\sim1/3$ of the total stellar mass of the Milky Way \citep{zoccali_3d_2016}; under the assumption that this is also true of the white dwarf binary population, we set $q_{\rm BD}=0.33$. 
Finally, we assume the GBs lie in an exponential disk with radial scale height of 3.31\,kpc per the thick disk Galaxy model of \citet{mcmillan_mass_2011}. 
These values are also listed in Table~\ref{tab:hyperpriors_truevals}.

We coarse-grain to a frequency resolution of $\Delta f=5\times10^{-5}$\,Hz for both simulation and computation of $\Sht$ and $\Nht$. 
This reduced frequency resolution is solely a matter of current computational limitations. 
The expense of the thresholding algorithm is largely driven by the number of frequency bins used and further optimization (or machine learning emulation) will be required to reach higher frequency resolutions; see further discussion on this point in \S\ref{sec:discussion}. 

We assume log-Normal distributed uncertainty on the total simulated PSD in each frequency bin, with a mean at the simulated sum of the foreground and noise PSDs and a conservative 0.2 dex standard deviation \citep[see e.g., Galactic foreground PSD uncertainties in][all of which assume simple models for the LISA instrumental noise]{karnesis_characterization_2021,pozzoli_cyclostationary_2024,criswell_templated_2025}. The final simulated spectrum is then produced by drawing these errors in every frequency bin and adding them to the true total PSD.

We use {\tt Eryn} \citep{karnesis_eryn_2023} to sample the population posterior. Therein, the full likelihood for the toy model is computed as follows. We draw values of $\{\mu_m,\sigma_m,\alpha_a,r_{\rm bulge},r_{\rm disk},q_{\rm BD}\}$ from their hyperpriors and condition the population priors accordingly. $N_r=5$ Galaxy realizations are drawn from the population model per each of 5 walkers and 15 parallel tempering temperatures, for a total of 375 simulated galaxies of 1,000,000 GBs each per likelihood evaluation. The thresholding algorithm is applied in parallel to all realizations, walkers, and temperatures to produce $N_r$ sets of $\Sht$ and $\Nht$. We compute $p(\Nres|\{\Nht\}_r)$ and $p(\Sgw|\{\Sht\}_r)$ via Eqs.~\eqref{eq:marginal_negative_binomial} and ~\eqref{eq:marginal_t}. We follow our simulation and take $\mathcal{L}(d|\Sgw)$ to be log-Normal with a mean of the simulated, error-perturbed spectrum and a standard deviation of 0.2 dex. We then convolve $\mathcal{L}(d|\Sgw)$ and $p(\Sgw|\{\Sht\}_r)$ over a grid in $\Sgw$ to calculate the integral over $\Sgw$ in Eq.~\eqref{eq:full_marginal_posterior}. We then compute the probability of the resolved binary parameters under the population-informed prior as given by Eq.~\eqref{eq:sam_pthetas_practical}.\footnote{ Note that no additional treatment of $p(d|\Nres)$ is applied, as we are assuming evaluation of a single block. In a Global Fit setting, however, this would be handled naturally by the changing number of resolved GBs at each state of the reversible-jump sampler as passed to the population module.} The sum of these log likelihoods is then the full likelihood of the toy model.

\begin{figure*}
    \centering
    \includegraphics[width=1\linewidth]{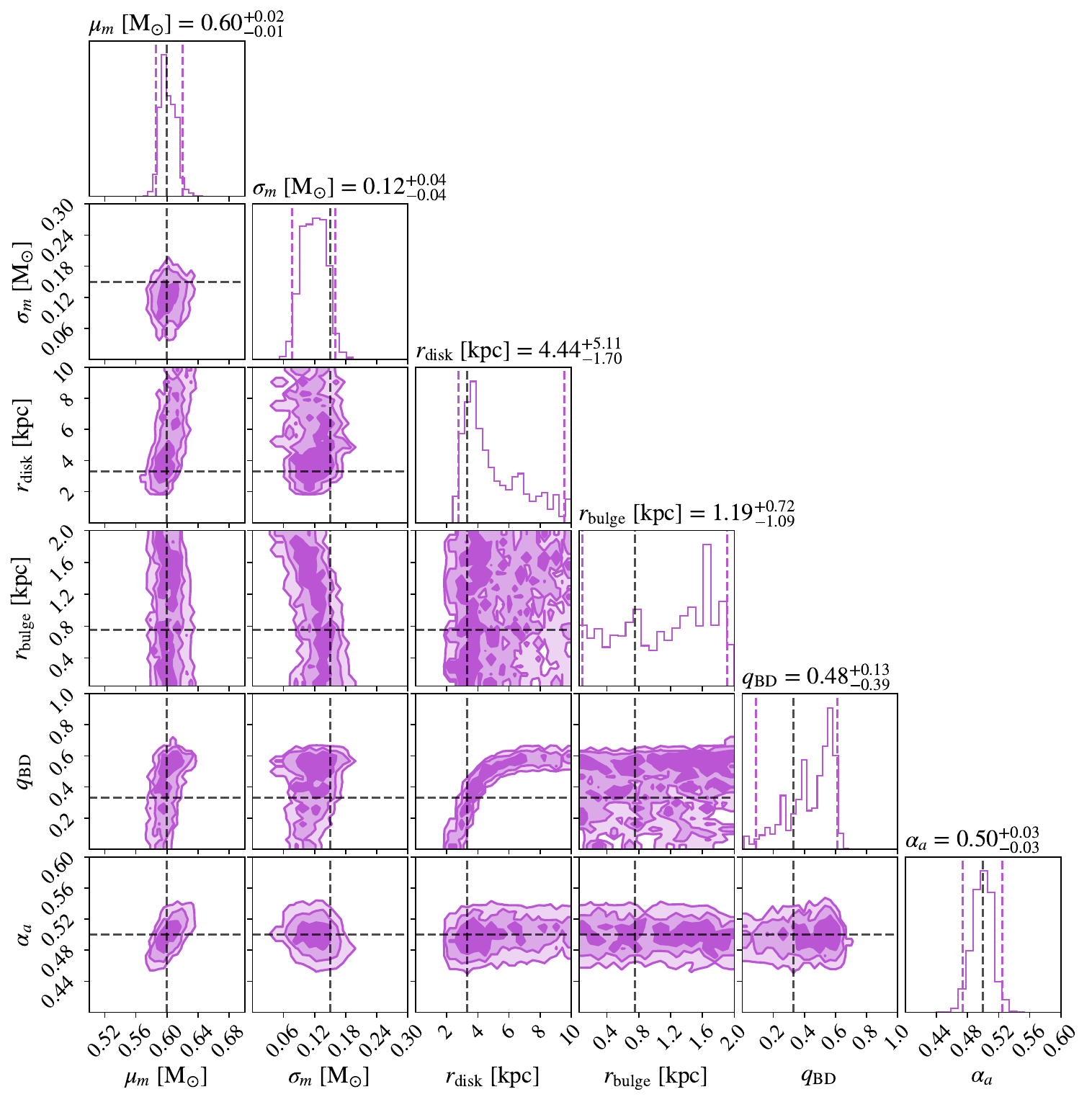}
    \caption{Corner plot showing the hyperparameter posteriors from the toy model population analysis. Contours shown are the 68\%, 95\%, and 99.7\% credible levels. The quoted constraints are the means and 95\% credible intervals. The dashed lines indicate the simulated values. We include all walkers of the zero-temperature chain and remove the first 2,500 samples from each walker's chain to account for burn-in.}
    \label{fig:popcorner}
\end{figure*}

\section{Results}\label{sec:Results}
We are able to successfully recover the parameters of our population model via the joint population inference formalism in the toy model setting of an \textit{in situ} global fit analysis.
A corner plot showing the population recovery can be found in Fig.~\ref{fig:popcorner}. 
In all cases, the simulated population hyperparameter values are recovered to within the 95\% credible region. 
Both the mass parameters and the orbital separation slope are well-recovered. It is interesting to note that there appears to be a slight covariance between the mass distribution mean $\mu_m$ and the orbital separation power law slope $\alpha_a$, presumably due to their shared contribution to the overall shape of the foreground spectrum.

While overall accurately recovered, the Galaxy model  has some difficulty ascertaining whether power is being contributed from the bulge or disk. 
In general the bulge characteristic radius $r_{\rm bulge}$ is not constrained, and we see a strong degeneracy between the disk scale height $r_{\rm disk}$ and the bulge-disk mass ratio $q_{\rm BD}$. 
This is likely due to two factors. 
First, our choice of a flattened 1D distance model; one would expect that accounting for the full 3D GB localizations and the Galactic foreground anisotropy will break the degeneracies we see here. 
Additionally, the parameterization of the exponential disk model allows for coupling between the number density in the bulge and the disk scale height. The disk exponential extends into the Galactic center; as such,  in cases where the bulge dominates, the disk is pushed to broader, flatter distributions which have reduced density in the regions where the bulk of the GBs reside. 
More detailed Galactic and stellar models will be considered in future work. That being said, despite the appearance of the 1D marginal distributions in $r_{\rm disk}$ and $q_{\rm BD}$, the 2D marginal posterior does indeed peak near these parameters' simulated values. 

\begin{figure*}
    \centering
    \includegraphics[width=1\linewidth]{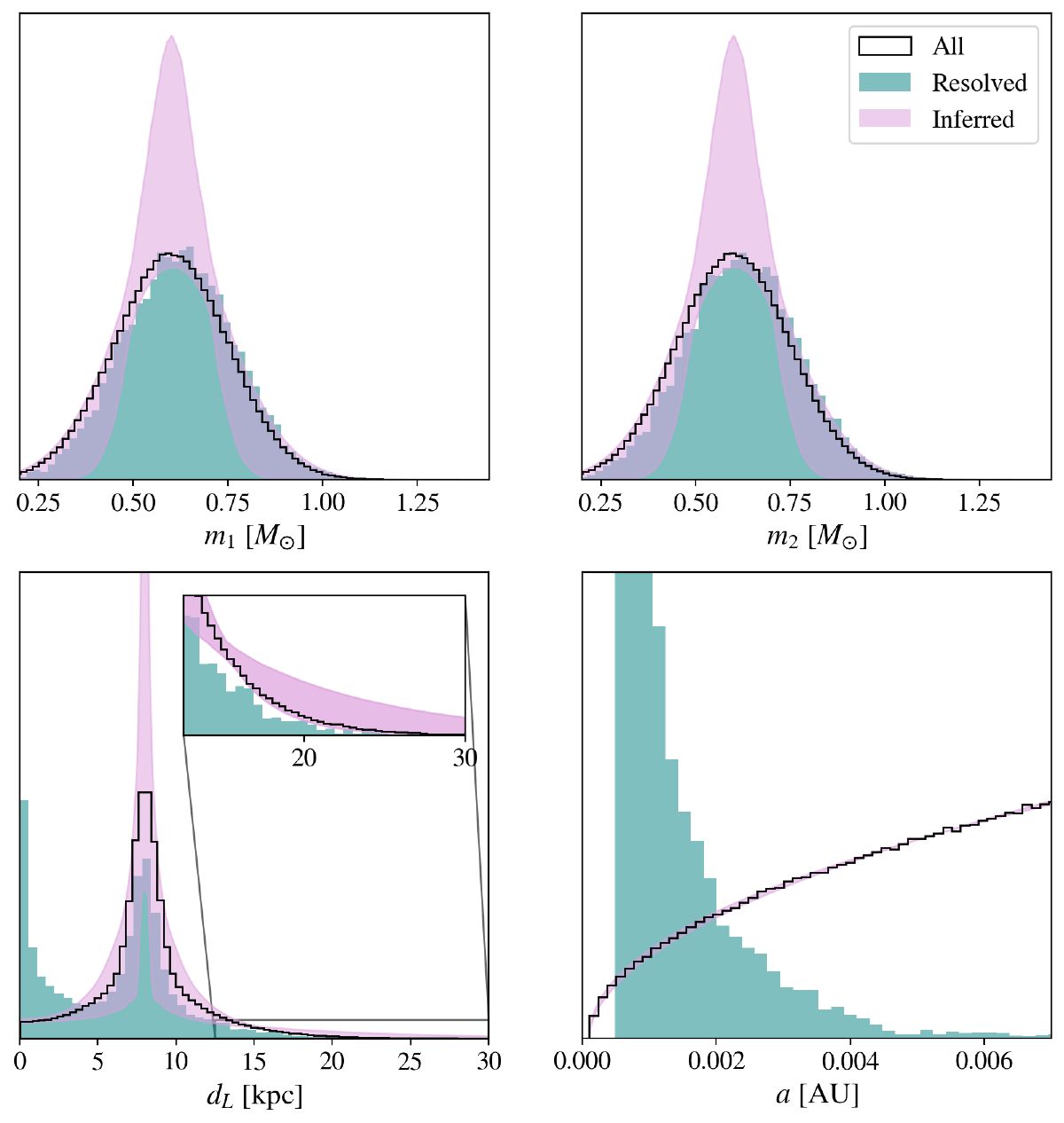}
    \caption{Recovered population distributions, plotted against the (density-normalized) histograms of both the full simulated GB population and the resolved GBs from that population. The inferred distributions are shown as 95\% credible intervals. The recovered distributions follow the full population. The differences between the resolved and unresolved subpopulations can be clearly seen; as would be expected, the resolved systems have higher masses and frequencies but sit at lower distances than their unresolved counterparts. The inferred distance distribution is strongly peaked at the Galactic center due to $r_{\rm bulge}$ being unconstrained; we cut off the top of the distribution for improved visibility of the histograms.}
    \label{fig:astro-dists}
\end{figure*}

Confirmation that the joint inference formalism is accurately recovering the distribution of the full population can be found in Fig.~\ref{fig:astro-dists}, which shows the recovered population distributions plotted against the histograms of the full simulated population and its subset of resolved GBs. Crucially, the distribution of the resolved GBs is starkly different to that of the full population. The resolved GBs sit on average at higher masses and smaller orbital separations in comparison to the unresolved GBs, as would be expected. Their distance distribution is bimodal, in part following the overall distribution of the Galaxy model with a peak at the Galactic bulge, but with another, stronger peak at low $d_L$. Additionally, the resolved GBs are underrepresented at large distances. These effects are expected due to the coupling of $d_L$ with the GW amplitude of the source. These trends are broadly similar to the resolvable and unresolvable populations estimated in \citet{Buscicchio:2024wwm}.

Despite the differences between the resolved GBs and the overall population, the inferred distributions clearly trace the full population; this can be seen most clearly in the distributions on $a$ and $d_L$. The mass distribution mean is recovered precisely at $\mu_m = 0.6^{+0.02}_{-0.01}\,\msun$, in agreement with the simulated value of $0.6\msun$, rather than the sightly higher mean of the resolved GB distribution seen in Fig.~\ref{fig:astro-dists}. We can conclude that the inferred mass distribution is also recovering the distribution of the full population. 


\begin{figure}
    \centering
    \includegraphics[width=1\linewidth]{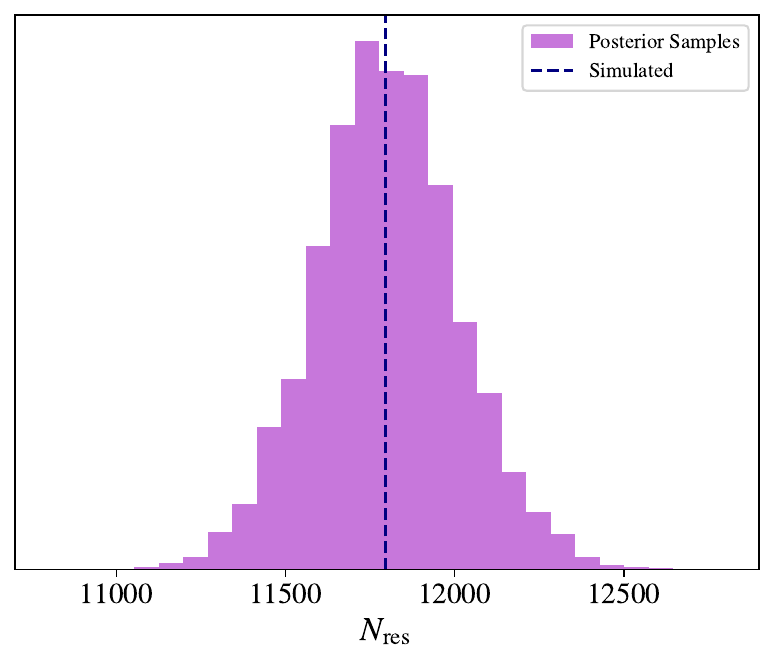}
    \caption{The population-derived prior on the number of resolved binaries.}
    \label{fig:nres_hist}
\end{figure}

\begin{figure*}
    \centering
    \includegraphics[width=0.8\linewidth]{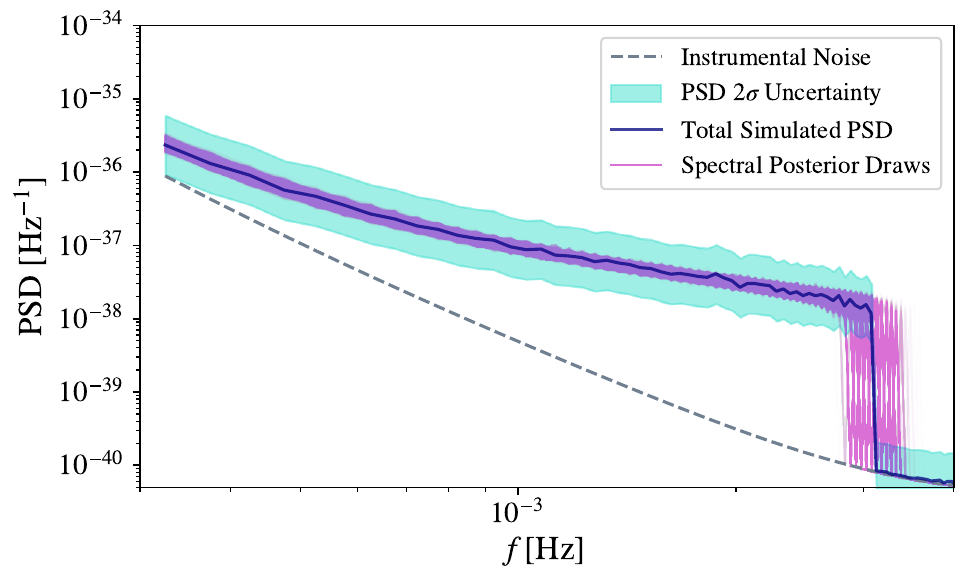}
    \caption{Posterior distribution of the total PSD spectrum, i.e. the sum of the instrumental noise and Galactic foreground contribution, for the toy model population analysis. The population-informed prior on the spectrum is strongly-informative without biasing the spectral posterior. While in principle a by-product of the population inference formalism, this indicates potential for this approach to produce a population-driven, astrophysically informed spectral model for the LISA Galactic foreground.}
    \label{fig:spectral_posterior}
\end{figure*}

We show in Fig.~\ref{fig:nres_hist} the latent distribution on the overall number of resolved binaries, which agrees with the simulated value of $\Nres$. In a realistic global fit analysis, this distribution provides a population-informed prior on the number of resolvable GBs. The population-informed recovery of the Galactic foreground spectrum is shown in Fig.~\ref{fig:spectral_posterior}. 
This latent posterior distribution is nominally a by-product of the population inference process. However, it shows intriguing potential as a direct model of the LISA Galactic foreground. The example shown here is able to naturally recreate the spectral shape of the Galactic foreground by directly modelling the precise astrophysical processes which give rise to that very shape. Moreover, it does so in a fashion directly guided by information from the resolved GBs and the overall population model. This results in much tighter posterior constraints than could be expected in isolation; c.f. the 2 sigma contours of the simulated PSD to the spread of the inferred spectral distribution.


We note that the spectral posterior in Fig.~\ref{fig:spectral_posterior} struggles to precisely characterize sharp features such as the rapid truncation around 3 mHz (which is itself an artifact of the coarse frequency resolution of the toy model), only constraining it to within $\sim10$ frequency bins ($3.1\pm0.1$ mHz). 
This is because we infer in this case the posterior of the full spectrum jointly across all frequency bins; variations in one frequency bin are therefore subject to the population draw as distributed across all frequency bins.

\section{Towards an Astrophysically Motivated Global Fit}\label{sec:vision}
The statistical framework and prototype analysis package presented in this work provide a foundation for the development of a fully astrophysically-motivated LISA global fit. That being said, several major challenges remain between the toy model analysis developed here and realistic implementation within a global fit.

\subsection{Astrophysical Realism}
Firstly, the population-driven spectral model for the Galactic foreground must be connected to a framework for LISA stochastic analyses on TDI data such as the Bayesian LISA Inference Package \citep[{\tt BLIP};][]{banagiri_mapping_2021,criswell_flexible_2025}. Doing so will not only allow for testing, validation, and optimization of such a foreground model on more realistic simulated data, but is also a crucial step towards applying the analysis to more complex datasets like LISA Data Challenge 2a ``Sangria'' \citep{lejeune_lisa_2022} or the upcoming Common Dataset 1 (``Mojito''). {\tt BLIP} in particular provides a robust suite for anisotropic modelling of stochastic GW signals in LISA data \citep[see][]{banagiri_mapping_2021,criswell_flexible_2025,criswell_templated_2025}, which will allow us to directly connect the GB population model to the anisotropy-induced temporal modulation of the Galactic foreground. This is not a trivial step; direct inference of astrophysical models of the Galactic foreground anisotropy is a computationally challenging problem. While the static anisotropic Galactic spatial templates of \citet{criswell_templated_2025} or the cyclostationary methods of \citet{pozzoli_cyclostationary_2024} provide potential prescriptive approaches, we would ideally be able to dynamically infer the Galactic foreground anisotropy via parameterized astrophysical models of the Milky Way. Work within {\tt BLIP} towards this end for the case of simple analytic spatial models of the Galaxy is ongoing (A. W. Criswell et al., in preparation). When that effort is complete, it will be relatively straightforward to tie in a population-driven model of the foreground anisotropy. Such an approach will allow the joint population inference to guide both spectral and spatial characterization of the Galactic foreground, further leveraging information from the resolved binaries to improve our modelling of the foreground.

Within the global fit, an embedded population model would be beneficial for the GB block, providing it with proposal guidance vis-\`a-vis population-informed priors on not only the overall number of resolved binaries as a function of amplitude and frequency, but also the number of subthreshold sources which contribute to residual non-Gaussianity \citep{rosati_prototype_2024, Buscicchio:2024wwm}. In fact, the framework presented here could be extended to encompass not just fully-stochastic and individually resolved sources, but also marginalization over all possible realizations of subthreshold GBs and their contribution to residual non-Gaussianity, thereby providing a useful handle on this potentially-problematic predicament. This can be accomplished by performing the thresholding process demonstrated here with realistic GB waveforms via codes like {\tt GBGPU} \citep{katz_bayesian_2022,katz_mikekatz04_2024}.

We do not currently consider the overlap between individual GBs and how this contributes to their confusion. \citet{Buscicchio:2024wwm} found the overlap to be strongest in the highest frequency area of the confusion noise where the unresolvable sources are smaller in number ($\sim 3\,\mathrm{mHz}$), so including these effects is likely to improve the PSD in this region.

To account for these effects, we will need to move beyond the simplifying choice made in the toy model of fixing $\rhothresh$. Instead, $\rhothresh$ should be parameterized and inferred by the data. This requires little extension of the formalism as presented, save that a prior on $\rhothresh$ be defined. Indeed, one need not condition on the SNR specifically; alternate formulations in terms of the GW amplitude or another characteristic quantity of interest can be constructed in the same manner. In the case of transdimensional analyses like the LISA global fit, the uncertainty on $\rhothresh$ should be naturally marginalized over as the set of systems considered resolved will vary jointly with the noise and Galactic foreground models. This uncertainty would then be correctly propagated into the uncertainty of our inference on the population parameters $\{N,\Lambda\}$.

Moving forward, several other aspects of realistic global fit GB analyses need to be taken into account. The toy model presented in this work considers only the astrophysical parameter space of interest. However, actual GB parameter inference \citep[e.g.,][]{cornish_tests_2007, katz_bayesian_2022, robson_detecting_2018} operates in a phenomenological parameter space, with $\vec\theta_{\rm phenom}\equiv \{A,f,\dot{f},\ddot{f},i,\phi_0,\psi,\hat\Omega\}$, i.e. the GW amplitude, frequency, frequency derivatives, inclination, initial phase, polarization angle, and sky position, respectively. It will not be trivial to transform distributions inferred within the astrophysical parameter space to the corresponding distributions within the phenomenological space.
Additionally, the presence of interacting systems within the population can give rise to ``pathological'' values of $\dot{f}$ and $\ddot{f}$ which are negative or otherwise deviate from the expectations for purely GW-driven evolution but are entirely allowed by the GB sampler. This at least can be addressed in a reasonably straightforward manner by decomposing (e.g.) $\dot{f}$ into its GW-driven evolution and evolution due to other astrophysical effects such that $\dot{f}=\dot{f}_{\rm GW} + \dot{f}_{\rm astro}$ and additionally inferring the distribution across the population of $\dot{f}_{\rm astro}$ (although more complex modelling choices could be made). For cases in which a significant portion of the population is eccentric, transient, and/or chaotic (e.g.,  EMRIs), an extension of the population inference formalism will need to be derived which accounts for frequency correlations.

On a final note with regard to astrophysical realism, the GB population will not solely consist of white dwarf binaries. It will instead be a heterogeneous mixture of white dwarf-, neutron star-, and black-hole-containing compact binary systems. With the rare exception of observably chirping systems, LISA will be unable to distinguish between these cases in the context of both resolved and unresolved GBs due to degeneracy between mass and distance in the GW amplitude. If this facet of the LISA GB population is left unaccounted for, our resulting population inferences will be biased as a result. Further complications arise given the fact that these populations will jointly form the Galactic foreground. As such, the set of resolved and unresolved systems for each subpopulation will depend not just on the LISA instrumental noise and the characteristics of that subpopulation, but also the characteristics of its fellow GB subpopulations, all layered atop the population-dependent circularity considered in this work. This is a challenging problem embedded within an already challenging problem, and implementation of heterogeneous population inference will need to wait until a successful proof-of-concept is established for the ``simple'' case of homogeneous GB population inference within the global fit. For the moment, however, we present an extension of the formalism presented in this work to the heterogeneous case of multiple, jointly-dependent subpopulations in Appendix~\ref{appendix:multipop}. 

\subsection{Practical Challenges}
Further difficulty lies in scaling the analysis presented here to a realistic setting. The frequency resolution currently attainable within \pelargir while maintaining reasonable sampling speed does not reach the level of the resolution used within GB analyses. When available GPU RAM can be solely allocated to frequency-parallelization, \pelargirs thresholding algorithm can reach reasonable speeds for realistic frequency resolutions. In practice, however, we require parallel tempering and multiple semi-analytic model realizations per likelihood evaluation. While the parallelization implemented within \pelargir is helpful, the nature of the thresholding algorithm is such that the arrays involved must have dynamic shapes. This aspect of the problem enforces a number of serial set-up operations which scale with the number of nominally-parallel evaluations and prevent use of potential efficiency improvements from compilation (either via Cupy's kernel caching or JAX's just-in-time compilation).\footnote{Incidentally, this dynamic array sizing is the reason \pelargir is written in Cupy instead of JAX.}

A potential omnibus solution to several of these challenges lies within normalizing-flow-based emulation. Such a procedure as applied to emulating semi-analytic models of GW source populations and their resulting GW background PSD has been demonstrated in pulsar timing arrays \citep{laal_deep_2025}. The thresholding module within \pelargir provides  an efficient route to create a training dataset for any semi-analytic model of the GB population which --- and this is vital --- includes the joint formalism developed in this work. The flow can then directly learn the probabilistic mapping, including the population-dependent division of resolved and unresolved binaries, of $\pi(\Sgw,\Nres,\thetasi|N,\Lambda,\eta)$. Additionally, by casting the astrophysical parameters of $\thetasi$ to the phenomenological space of $\vec\theta_{\rm phenom}$ in the training data, the flow can instead learn $\pi(\Sgw,\Nres,\{\vec\theta_{\mathrm{phenom},i}\}|N,\Lambda,\eta)$. This provides an elegant solution to the issue of inferring astrophysical distributions from phenomenological GB models. Training a flow to map from an $\mathcal{O}(5-10)$-dimensional astrophysical parameter space to $\mathcal{O}(10^5)$-dimensional space (the number of frequency bins in the relevant part of the LISA frequency band) is a highly non-trivial proposition. However, by training a separate flow for each frequency bin (or set of several adjacent frequency bins), we not only overcome the dimensionality problem but also allow for the creation of a model which can directly infer the number of binaries (both resolved and unresolved), the foreground amplitude, and the expected characteristics of resolved binaries within each frequency bin. Such a per-frequency structure mirrors the practical construction of the GB blocks --- search and parameter estimation operations in parallel across frequency bins --- and can provide population-informed priors for both GB parameter estimation and proposals to add or remove new resolved sources.

Care will need to be taken in the implementation of this kind of approach within a proper global fit environment. Both GLASS \citep{littenberg_prototype_2023} and Erebor \citep{katz_efficient_2024} use half-overlapping frequency bins to account for correlations between adjacent bins; this will need to be accounted for statistically to avoid double-counting resolved sources. This kind of approach will also greatly improve the efficiency of the population inference analysis. The thresholding algorithm is the most expensive part of the \pelargir population likelihood as it stands; flow emulation should allow for $\sim$sub-millisecond population-level likelihood evaluations, leaving the anisotropic stochastic likelihood and the GB sampler as the main bottlenecks. Finally, population inference in the global fit will be most effective following the initial search phase, once the fit has quickly burned in to a reasonable set of initial resolved GBs \citep[see][for discussion of initial search phases in the global fit]{littenberg_prototype_2023,katz_efficient_2024,deng_modular_2025}.

\section{Discussion}\label{sec:discussion}
We present a holistic population-modelling formalism for joint inference of a set of individually-resolved GW sources and an unresolved astrophysical foreground. 
We additionally present \pelargirx, a prototype GPU-accelerated global fit module which implements the joint population inference framework and has been designed for compatibility with current global fit prototypes. Included within \pelargir is the thresholding module, which can perform rapid evaluation of the joint population framework presented in this work. The thresholding algorithm is fast ($\mathcal{O}$(0.1 s) evaluation time for the simple cases considered in \S\ref{sec:toymodel}, and $\mathcal{O}$(6 s) for full population synthesis catalogues), CPU/GPU agnostic, and has broad applicability outside of a pure inference setting for estimating which members of a simulated GB population will be resolved or unresolved. We demonstrate the formalism and \pelargir with a toy model of the LISA white dwarf binary population, and show that the framework can successfully recover hyperparameters describing the full simulated population, its total number of resolved binaries, and the simulated Galactic foreground PSD. Finally, we lay out a roadmap for the creation of an astrophysically-motivated LISA global fit with embedded population inference.

While we focus in this work on the case of the LISA GB population, we stress that this formalism is applicable across the gravitational-wave spectrum. When pulsar timing arrays begin to resolve individual SMBHBs from the nHz GW background, the framework presented here will allow for joint inference of the SMBHB population while inherently accounting for the impact of the nHz GW background on individual SMBHB detection, the power removed from the GW background by each given proposed individual system, as well as the Poisson shot noise expected to dominate the first few resolved SMBHBs. Moreover, given current non-detection of individual SMBHBs \citep[e.g.,][]{agazie_nanograv_2023e}, a special case of this formalism with zero resolved systems can be applied to pulsar timing datasets today. In next-generation terrestrial observatories like Cosmic Explorer and Einstein Telescope, confusion noise from binary neutron star mergers is expected to impact individual event characterization; an extension of this framework to transient sources will likely be required for population inference in that setting. Depending on the number of EMRIs present in LISA data, a related approach may also be required for inference of the EMRI population; the joint dependence of the resolved GB and EMRI populations on their respective astrophysical foregrounds would necessitate implementation of the extension of this formalism to multiple coexisting source populations presented in Appendix~\ref{appendix:multipop}.

In addition to the vision laid out in \S\ref{sec:vision} for an astrophysically-motivated LISA global fit, there exist a number of interesting directions of future research which build on this work. First, while here we construct a simplified model of the GB population, the formalism can in principle handle any level of model complexity. More advanced semi-analytic models can be straightforwardly implemented within \pelargir to explore the potential of this approach for inference of more involved details of Galactic and/or stellar astrophysics. In principle, the normalizing-flow-based implementation discussed in \S\ref{sec:vision} will even allow for entire GB population synthesis models to be embedded --- and compared --- via emulation within the LISA global fit. Generating the training data for such an application would be an intensive process, but flow emulation would amortize this cost and allow for rapid evaluation of the underlying astrophysics against the observed GB population. 

Additionally, there will be likely be GW background contributions \citep{rieck_stochastic_2024,pozzoli_cyclostationary_2024} and resolved binaries \citep{korol_populations_2020} in LISA data stemming from Milky Way satellite galaxies like the Large Magellanic Cloud (LMC). This provides an interesting situation in which the ability to resolve the Milky Way's GBs is largely independent of the LMC population, whereas our ability to resolve individual systems in the LMC is absolutely dependent on the nature of the GB population. Adding to this is the fact that we will not \textit{a priori} know whether a given binary resides in the Milky Way or the LMC. The framework presented in this work can also allow for joint modelling of the Galactic and (e.g.) LMC mHz compact binary populations, providing a new window into the stellar and dwarf-galactic astrophysics of our closest neighbors.

Finally, we return to the contrast of in-post vs. \textit{in situ} approaches for inference of the LISA GB population. For the reasons discussed in \S\ref{sec:intro}, there is a concern that in-post population inference on the output of the LISA global fit will struggle with issues of resampling efficiency and accuracy, possibly biasing the results of such analyses in the case of incomplete global fit convergence. However, the extent of these issues is at present unclear. In order to ascertain the impact and severity of any potential barriers, both the approach proposed in this work and the complementary, in-post method of \citet{toubiana_framework_2026} should be developed to an advanced state, at which point the results of both methods can be compared as applied to a realistic common dataset with known underlying physics.

From a technical perspective, it may be possible to circumvent some (if not all) of the issues raised in \S\ref{sec:intro} by embedding maximally-flexible, purely phenomenological population models within the global fit. The result of such an approach would be posteriors on the distributions of GW amplitudes, frequencies, etc. of the \textit{full} population, with circularity and other complications already accounted for. This kind of data product could allow the broader astronomical community to fit their models to these distributions rather than the global fit posterior itself. Provided that the population models used in the original global fit analysis are sufficiently non-prescriptive that the posterior density for any downstream model of interest would be fully encompassed by the posterior space of the original phenomenological inference, such an approach could in principle allow for unbiased in-post inference of astrophysical quantities.

Independent of whether or not its existence is necessary for (approximately) unbiased population inference with LISA, an astrophysically-motivated global fit with embedded population inference has the potential to provide numerous benefits for LISA data analysis. It produces population-informed priors on the resolved GBs, which even beyond improving parameter inference will likely result in more, more confident resolved GBs \citep{littenberg_prototype_2023}. Moreover, it has the potential to allow for population-guided priors on not just the number of resolved binaries in each frequency bin, but also the distribution --- and therefore induced non-Gaussianity --- of the set of currently subthreshold systems. Furthermore, joint inference of the resolved and unresolved GB population provides a natural, astrophysical model for the LISA Galactic foreground, which can then be directly informed in terms of both its spectrum and its anisotropy by the properties of the resolved binaries. This approach has the potential to significantly improve our ability to accurately characterize the Galactic foreground and therefore --- due to the latter's status as a significant source of astrophysical noise for most other LISA analyses --- improve almost every LISA science case.

\section*{Acknowledgments}
The authors would like to thank Pat Meyers, Jacob Golomb, Asad Hussain, and Katerina Chatziioannou for their contributions to technical discussions on the problem of population inference with LISA at the 2024 LISA Sprint. The authors would also like to thank Katerina Chatziioannou, Tyson Littenberg, Michael Katz, Alexandre Toubiana, and Astrid Lamberts for helpful discussions in the time since. This project was initiated at the 2024 LISA Sprint,
which was hosted by Caltech and supported by the Jet
Propulsion Laboratory Astronomy and Physics Directorate. AWC acknowledges support by NSF NRT-2125764. SRT acknowledges support from NSF AST2307719, an NSF CAREER PHY-2146016, and a Vanderbilt University Chancellor's Faculty Fellowship. AWC and SRT acknowledge support from NASA LPS-80NSSC26K0342. S.B. acknowledges support from Australian Research Council grants CE230100016, LE210100002, DP230103088 and LE260100008, and from NSF grant PHY-2207945 (PI: Kalogera).

\bibliography{zotero_refs,additional_refs}{}
\bibliographystyle{aasjournalv7}


\appendix

\section{Derivation of Marginal Poisson Term}\label{appendix:formalism_marginal_poisson}
\begin{figure}
    \centering
    \includegraphics[width=1\linewidth]{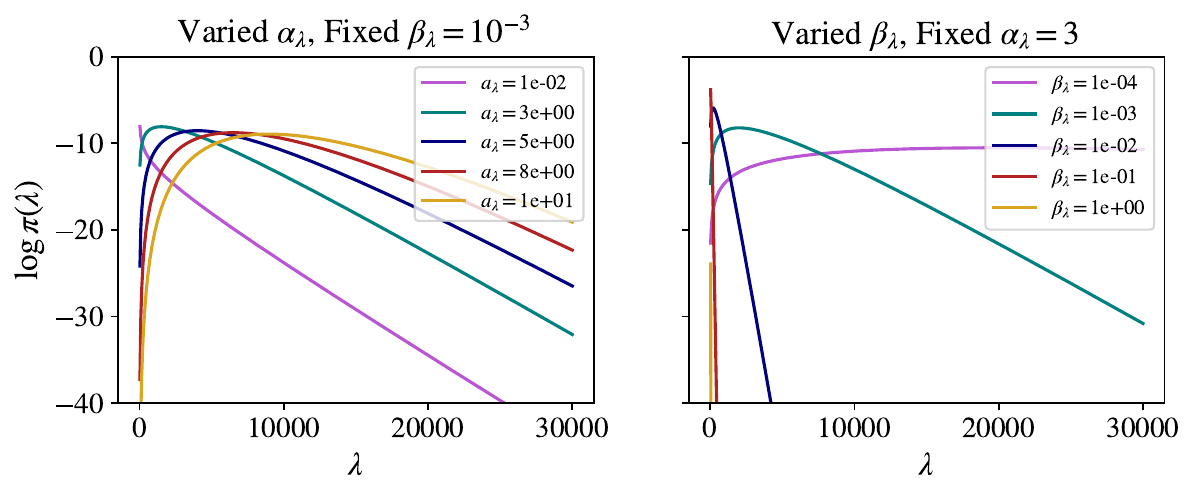}
    \caption{Log prior distributions on $\lambda$ for different choices of the hyperprior values.}
    \label{fig:conjugate-priors-lambda}
\end{figure}

We require the probability $p(\Nres|\Nht)$ that the observed $\Nres$ and the semi-analytic-model-evaluated $\Nht$ are Poisson realizations from the same underlying rate $\lambda$, marginalized over our uncertainty as to that rate.
In practice, we consider some finite number $N_r$ realizations of the Galaxy model, each with its own $\Nht$. 
Collectively, let this set of realizations be denoted $\{\Nht\}_r$; we are therefore interested in the following:
\begin{equation}
    p(\Nres|\{\Nht\}_r) = \int p(\Nres|\lambda)p(\lambda|\{\Nht\}_r)d\lambda\,.
    \label{eq:rate-marg}
\end{equation}
The first term in the integrand is trivial: 
\begin{equation}
    p(\Nres|\lambda) \sim \mathrm{Poisson}(\Nres|\lambda)\,.
\end{equation}
If we choose a conjugate prior on $\lambda$ such that
\begin{equation}
    \pi(\lambda) \sim \mathrm{Gamma}(\alpha_{\lambda},\beta_{\lambda})\,,
\end{equation}
we can use this prior and Bayes' theorem to write
\begin{equation}
    p(\lambda|\{\Nht\}_r) = \frac{p(\{\Nht\}_r|\lambda)\pi(\lambda)}{p(\{\Nht\}_r)} \propto p(\{\Nht\}_r|\lambda)\pi(\lambda)\,,
\end{equation}
where for each realization
\begin{equation}
    p(\Nht|\lambda)\sim \mathrm{Poisson}(\Nht|\lambda)\,.
\end{equation}
Due to our choice of a conjugate Gamma prior for $\lambda$, $p(\lambda|\{\Nht\}_r)$ is also a Gamma distribution such that
\begin{equation}\label{eq:conjugate_gamma}
    p(\lambda|\{\Nht\}_r)\propto \mathrm{Gamma}\left(\alpha_{\lambda}'=\alpha_{\lambda}+\sum_{i=1}^{N_r} \hat{N}_{\mathrm{res},i},\beta_{\lambda}' = \beta_{\lambda}+N_r\right).
\end{equation}
Returning to Eq.~\eqref{eq:rate-marg}, the full expression is now
\begin{equation}
    p(\Nres|\{\Nht\}_r) \propto \int\,\mathrm{Poisson}(\Nres|\lambda) \times \mathrm{Gamma}\left(\lambda|\alpha_{\lambda}',\beta_{\lambda}'\right)\,d\lambda\,.
\end{equation}
This is a mixed Poisson-Gamma distribution, which is equivalent to a negative binomial distribution \citep{willmot_mixed_1986}. 
Therefore, the analytic marginal distribution for the observed number of resolved binaries $\Nres$ given some $N_r$ realizations of the population-model-predicted $\Nht$ is:
\begin{equation}\label{eq:marginal_negative_binomial}
    p(\Nres|\{\Nht\}_r) \propto \mathrm{NegBin}\bigg( \Nres \Big|\, r_{\mathrm{NB}}=\alpha_{\lambda}',\ p_{\mathrm{NB}}=\frac{\beta_{\lambda}'}{1+\beta_{\lambda}'}\bigg)\,,
\end{equation}
where $r_{\mathrm{NB}}$ and $p_{\mathrm{NB}}$ are the standard $r$ and $p$ parameters of the Negative Binomial, and $\alpha_{\lambda}'$ and $\beta_{\lambda}'$ are as defined in Eq.~\eqref{eq:conjugate_gamma}. This result is valid for any positive integer $N_r$. The approach given here requires a choice of the Gamma distribution hyperparameters $\{\alpha_{\lambda}, \beta{\lambda}\}$, which then describe a prior on the total rate of (resolved) GBs. In general, the marginal distribution is robust to choice of $\{\alpha_{\lambda}, \beta_{\lambda}\}$ (i.e., the result is dominated by the values of $\{\Nht\}_r$), provided $\alpha_{\lambda}$ is of order unity and $\beta_{\lambda} \ll 1$. For the analyses presented in this work, we choose $\alpha_{\lambda}=3$ and $\beta_{\lambda}=10^{-3}$, which corresponds to a broad prior on the rate $\lambda$ (see Fig.~\ref{fig:conjugate-priors-lambda}).



\section{Derivation of Marginal Gaussian Term}\label{appendix:formalism_marginal_gauss}
In principle, the formalism in \S\ref{appendix:formalism_marginal_poisson} can be generalized to the fully stochastic case, as the total number of unresolved binaries is also a Poisson process. 
However, bridging between the discrete nature of Poisson statistics and the necessarily continuous space of the foreground PSD presents practical challenges in terms of computational representation.\footnote{One would in principle define a characteristic amplitude $\Bar{A}_i$ for the unresolved GBs in frequency bin $i$ such that $S_{\mathrm{GW},i} = \Bar{A}_i\times N_{\mathrm{unres},i}$, and then treat the Poisson statistics of $N_{\mathrm{unres},i}$ as laid out in \S\ref{appendix:formalism_marginal_poisson}. However, this has the side effect of discretizing the resulting marginal probability distribution $p(\Sgw|\Sht)$. As $\Sgw$ is a continuous variable on the reals, it becomes infinitely likely that a given value of $\Sgw$ has zero probability under $p(\Sgw|\Sht)$. This is actually less of an issue in a realistic global fit setting, where we would hierarchically sample $\Sgw$ from $p(\Sgw|\Sht)$ and evaluate it against the likelihood given in Eq.~\eqref{eq:whittle_likelihood}. However, for the toy model considered in this work which convolves $p(\Sgw|\Sht)$ with an abstracted analytic likelihood for $p(d|\Sgw)$ (see \S\ref{sec:toymodel}), this presents an intractable issue for definition of the integration grid.} We consider an alternate formulation that operates in the Gaussian limit of the Poisson distribution for a large number of systems. In our derivation, we consider a single frequency bin; but we assume that the foreground spectrum is statistically independent across frequency bins apart from the effects of the population, and as such this result can be trivially generalized to all frequency bins.

Via the central limit theorem, let $\Sgw$ and $\Sht$ be drawn from a Gaussian distribution with unknown mean $\mu_S$ and variance $\sigma_S^2$.
We will again estimate and marginalize over these unknown parameters via $N_r$ realizations of the population model spectra, denoted $\{\Sht\}_r$, such that:
\begin{equation}
\begin{split}
    p(\Sgw|\{\Sht\}_r) \propto \iint &p(\Sgw|\mu_S,\sigma_S)\\
    &\times p(\{\Sht\}_r|\mu_S,\sigma_S)\\
    &\times \pi(\mu_S,\sigma_S)\ d\mu_S d\sigma_S\,,
\end{split}
\end{equation}

where we have implicitly applied Bayes' theorem.
We will again choose a conjugate prior on $\{\mu_S,\sigma_S\}$, taking it to be a Normal-Inverse-Gamma distribution such that
\begin{equation}
    \pi(\mu_S,\sigma_S) \sim \mathcal{N}\mhyphen\Gamma^{-1}(\mu_S,\sigma_S|\mu_0,\nu,\alpha_S,\beta_S)\,,
\end{equation}
where $\{\mu_0,\nu,\alpha_S,\beta_S\}$ are prior hyperparameters to be chosen; $\mu_0$ is the initial guess as to the foreground amplitude, $\nu$ will be the weight of that guess with respect to the population model draws, and $\{\alpha_S,\beta_S\}$ determine the overall spread of the hyperprior. The corresponding posterior distribution under this conjugate prior is then also a Normal-Inverse-Gamma \citep{fink_compendium_1997}:
\begin{equation}
    p(\mu_S,\sigma_S|\{\Sht\}_r) \sim \mathcal{N}\mhyphen\Gamma^{-1}(\mu_S,\sigma_S|\mu_0',\nu',\alpha_S',\beta_S')\,,
\end{equation}
where
\begin{align}
    \mu_0' &= \frac{\nu\mu_0+N_r\Bar{S_r}}{\nu+N_r}\,,\\
    \nu' &= \nu+N_r\,,\\
    \alpha_S' &= \alpha_S'+\frac{N_r}{2}\,,\\
    \beta_S' &= \beta_S + \frac{1}{2}\sum_{i=1}^{N_r}(\hat{S}_{\mathrm{GW},i}-\Bar{S_r})^2 + \frac{\nu N_r}{\nu+N_r}\frac{(\Bar{S_r}-\mu_0)^2}{2}\,,
\end{align}
such that $\Bar{S_r}$ and $\hat{S}_{\mathrm{GW},i}$ denote the mean and elements of $\{\Sht\}_r$, respectively.

The integral over these terms will then be described by the (location-scale) t distribution \citep{denison_bayesian_2002}:
    \begin{equation}\label{eq:marginal_t}
\boxed{
    p(\Sgw|\{\Sht\}_r) \propto t_{2\alpha'_S}\left( \Sgw \Big| \mu_t=\mu_0', \sigma_t^2=\frac{\beta'_S(\nu'+1)}{\alpha'_S\nu'}\right)\,,}
\end{equation}
where the location-scale t distribution of a variable $x$ given location $\mu_t$, scale $\sigma_t^2$, and degrees of freedom $\nu_t$ is defined as:
\begin{equation}
    t_{\nu_t}(x|\mu_t,\sigma_t^2) = \frac{\Gamma(\frac{\nu_t+1}{2})}{\Gamma(\frac{\nu_t}{2})\sqrt{\pi\nu_t\sigma_t^2}} \left( 1+\frac{1}{\nu_t}\frac{(x-\mu_t)^2}{\sigma_t^2} \right)^\frac{-(\nu_t+1)}{2}\,.
\end{equation}


In the limit of $N_r\gg1$, this result converges to a Gaussian with a mean and variance equal to the mean PSD value across realizations, as expected from the limit of the underlying Poisson statistics. 
The heavy tails of the t distribution directly encode the model uncertainty we induce by considering a finite number of realizations.

\begin{figure}
    \centering
    \includegraphics[width=1\linewidth]{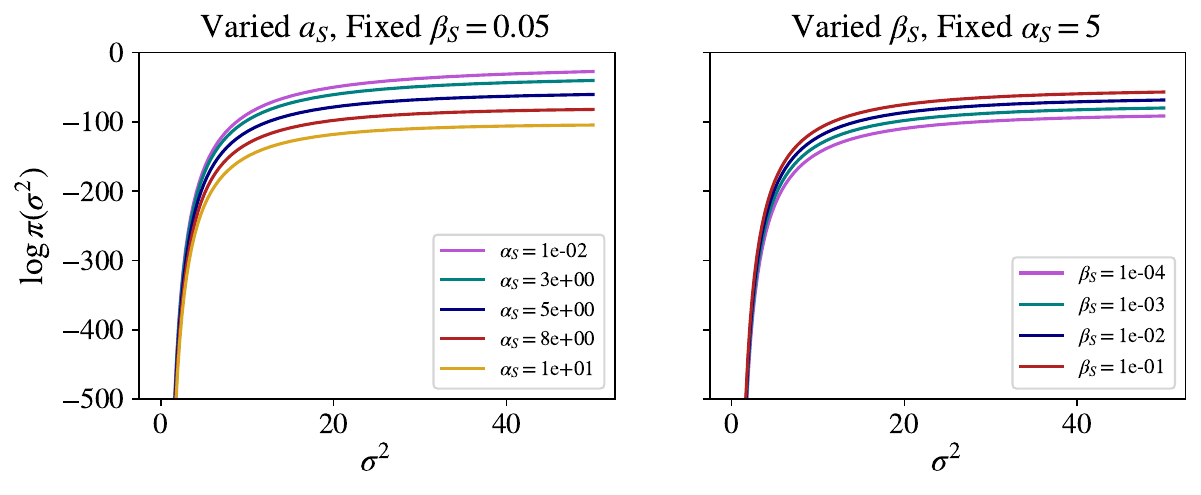}
    \caption{Log prior distributions on $\sigma^2$ for different choices of the hyperprior values.}
    \label{fig:conjugate-priors-sigma}
\end{figure}

As before, we must make a choice as to the values of our conjugate prior hyperparameters. 
In general, we find that the marginal distribution is robust under the following choices. 
We set $\nu\ll1$ so that the marginal distribution is largely insensitive to the initial prior mean $\mu_0$ provided it is comparable to typical values of $\log_{10}\Sgw$. As before, $\alpha_S$ should be of order unity. The final distribution is somewhat more sensitive to the value of $\beta_S$ as it governs the spread of the hyperprior; in general, we find our results are robust for values of $\beta_S$ comparable to the expected scatter of the foreground spectrum in dex across the population model.\footnote{That is, for a given spectrum, how much Poisson scatter is observed from frequency bin to frequency bin.} 
This approach is only valid for $N_r>1$; for a single realization the distribution in Eq.~\eqref{eq:marginal_t} has infinite variance. 
For the analyses presented in this work, we set $\nu=10^{-10}$, $\log_{10}\mu_0=-40$, $\alpha_S=5$, and $\beta_S=0.05$. This sets a marginal prior on $\sigma^2$ which is allowed to have arbitrarily large values but disfavors small variances. The impact of hyperparameter choice on $\sigma^2$ is shown in Fig.~\ref{fig:conjugate-priors-sigma}; $\nu$ is chosen to make the effect of the initial guess $\mu_0$ negligible, so we do not provide a corresponding plot on the mean.

\section{Extension to Multiple Overlapping Populations}\label{appendix:multipop}

In reality, the LISA GB population will not solely consist of white dwarf binaries, and will instead be comprised of some mix of white dwarf (WD), neutron star (NS), and black hole (BH) binaries, including both homogeneous and heterogeneous binaries (e.g. WD+WD vs. WD+BH). Inference of these contingent subpopulations will be necessarily deeply interconnected, as the circular population model considered in this work will be shared across subpopulations. For example, both the WD+WD and BH+BH subpopulations will have some number of resolved binaries and contribute to the Galactic foreground at some level; as a result, our ability to resolve individual BH+BH systems will be dependent on the level of the WD+WD foreground contribution, which will depend on our ability to resolve WD+WD systems, which will depend (in part) on the BH+BH foreground contribution, and so on. On top of this, it will not be possible in most cases to distinguish between members of different subpopulations as seen in the data. Unless the system is noticeably chirping, the nature of a given resolved GB will be muddied by the degeneracy in the GW amplitude $A$ between mass and distance. Systems contributing to the Galactic foreground will of course be completely indistinguishable on an individual basis. However, a population-driven approach gives us our best window on these distinct subpopulations. In connection with a 3D spatial model of the Milky Way, astrophysically-motivated distributions (and bounds) on the masses of a given subpopulation provide a probabilistic handle on this problem. For instance, a population-level model allows us to compare the relative probability of a massive BH+BH in the far reaches of the Galactic disk (lower) vs. a WD+WD in the Galactic bulge (higher); similarly, we can compare and account for the probability of an extremely close WD+WD (lower) vs. a farther off NS+NS (higher). This approach will not provide precise distinguishing power in most cases, but allows us to accurately apply astrophysical constraints, properly account for this uncertainty in our analysis, and make broader, population-level statements despite uncertainty at the level of individual systems.

To this end, we derive an extension of the formalism presented in \S\ref{sec:formalism} to the case of a population composed of an arbitrary number of subpopulations. Let the overall population be composed of $P$ subpopulations, such that the total number of systems across all subpopulations $N$ is now given as
\begin{equation}
    N = \sum_{k=1}^{P}N_P\,,
\end{equation}
where $N_P$ denotes the number of systems in population $P$ and is a quantity to be inferred. Furthermore, let
\begin{equation}
    \{N^*,\Lambda^*\}_P \equiv \big\{\{N_1,\Lambda_1\},\{N_2,\Lambda_2\},\ldots,\{N_P,\Lambda_P\}\big\}\,,
\end{equation}
and similarly,
\begin{equation}
    \{\Sgw^*,\Nres^*,\{\vec\theta_i^*\}\}_P \equiv \big\{\{S_{\mathrm{GW},1},N_{\mathrm{res},1},\{\vec\theta_{i,1}\}\},\{S_{\mathrm{GW},2},N_{\mathrm{res},2},\{\vec\theta_{i,2}\}\},\ldots,\{S_{\mathrm{GW},P},N_{\mathrm{res},P},\{\vec\theta_{i,P}\}\}\big\}\,,
\end{equation}
such that the total unresolved confusion noise $\Sgw$ is
\begin{equation}
    \Sgw = \sum_{k=1}^{P} S_{\mathrm{GW},k}\,,
\end{equation}
and the total number of resolved binares $\Nres$ is
\begin{equation}
    \Nres = \sum_{k=1}^{P} N_{\mathrm{res},k}\,.
\end{equation}
Finally, let $\Xi$ denote some set of shared underlying astrophysics which impacts all the subpopulations (e.g., the overall Galaxy model or the Galactic star formation rate). We can then write the equivalent of Eq.~\eqref{eq:full_posterior} for the multi-population case:
\begin{equation}\label{eq:full_multi_posterior}
    \begin{split}
    p(\{N^*,\Lambda^*\}_P,\Xi,\Sgw,\Nres,\thetasi,\eta|d) \propto&\ \mathcal{L}(d|\Sgw,\Nres,\thetasi,\eta)\,\pi(\Sgw,\Nres,\thetasi|\{N^*,\Lambda^*\}_P,\eta)\\
    &\qquad\qquad\qquad\qquad\qquad\qquad\qquad\qquad\times \pi(\eta)\,\pi(\{N^*,\Lambda^*\}_P|\Xi)\,\pi(\Xi)\,.
    \end{split}
\end{equation}
The equivalent of the marginal, population-informed prior in Eq.~\eqref{eq:sam_pthetas_practical} is then a mixture model for each system across subpopulations such that
\begin{equation}\label{eq:multi_marginal_prior}
    \pi(\thetasi|\{N^*,\Lambda^*\}_P,\Xi) = \sum_{k=1}^P \pi(\thetasi|N_k,\Lambda_k)\,\pi(N_k|\Xi)\,,
\end{equation}
with
\begin{equation}
    \pi(\thetasi|N_k,\Lambda_k) = \iiint \pi(\Sgw,\Nres,\thetasi|N_k,\Lambda_k,\eta)d\Sgw d\Nres d\eta\,,
\end{equation}
where the dependence of each subpopulation on the others is implicitly marginalized via the integrals over $\Sgw$ and $\Nres$. Note that in Eq.~\eqref{eq:multi_marginal_prior} we have explicitly included the relative subpopulation weights $\pi(N_k|\Xi)$ for clarity, whereas this dependence is implicit in Eq.~\eqref{eq:full_multi_posterior}. In the case of a system which is indeed distinguishable, i.e. a chirping system for which we can measure $\mathcal{M}$, this formulation is unchanged; some of the mixture model contributions to the sum in Eq.~\eqref{eq:multi_marginal_prior} will simply be zero or negligible for the bulk of the likelihood density for that system.

To compute $\pi(\Sgw,\Nres,\thetasi|\{N^*,\Lambda^*\}_P,\eta)$ in the case of a multi-component semi-analytic model, we can largely follow the prescription in \S\ref{sec:formalism_sams} sans modification. The marginal terms of $p(\Nres|\Nht)$ and $p(\Sgw|\Sht)$ are unaffected, as these terms both as observed and in the semi-analytic model will be in aggregate across all subpopulations. All systems, regardless of origin, should be ordered together from lowest to highest na\"ive SNR as discussed in \S\ref{sec:formalism_sams}. One can then compute $\{\Sht^*,\Nht^*\}_P$ in each frequency bin $j$ for each subpopulation $k$ with a slight modification to Eqs.~\eqref{eq:sam_nres_practical}--\eqref{eq:sam_sgw_practical}, taking into account the fact that we know the origin of each system within a draw from the semi-analytic model and can label them as such:
\begin{equation}\label{eq:multi_sam_nres_practical}
    \boxed{
    \hat{N}_{\mathrm{res},k}(f_j;\{N^*,\Lambda^*\}_P,\Xi,\eta,\rhothresh) = \sum_{i=1}^N \Theta\left\{\hat\rho_c(\vec\theta_i^{\,k'})  > \hat\rho_{\mathrm{boundary}} \right\}\delta(f_i-f_j)\delta(k-k')\,,
    }
\end{equation}
and
\begin{equation}\label{eq:multi_sam_sgw_practical}
    \boxed{
    \hat{S}_{\mathrm{GW},k}(f_j;\{N^*,\Lambda^*\}_P,\Xi,\eta,\rhothresh) = \sum_{i=1}^N S(\vec\theta_i^{\,k'})\Theta\left\{\hat\rho_c(\vec\theta_i^{\,k'}) \leq \hat\rho_{\mathrm{boundary}} \right\}\delta(f_i-f_j)\delta(k-k')\,,
    }
\end{equation}
where $\vec\theta_i^{\,k'}$ denotes the parameters of a binary drawn from subpopulation $k'$ with $k'\in\{1\ldots P\}$. The population-informed prior on all resolved binaries is then
\begin{equation}
\boxed{
\begin{split}
    p(\thetasi|\{N^*,\Lambda^*\}_P,\eta) &= \sum_{i=1}^{\Nres} \sum_{k=1}^P \left[\pi(\vec\theta_i|\Lambda_k)\,p(\mathrm{resolved}|\vec\theta_i,\{N^*,\Lambda^*\}_P,\eta)\right]\\
    &\simeq \sum_{i=1}^{\Nres} \sum_{k=1}^P\left[\pi(\vec\theta_i|\Lambda_k)\,\frac{1}{N_r}\sum_{\ell=1}^{N_r} \Theta\left\{ \rho\left(\vec\theta_i,\eta,\hat{S}_{\mathrm{GW},\ell}\right) \geq \rho_{\rm thresh} \right\}\right]\,.
\end{split}
}
\end{equation}

This approach naturally handles cases like the LMC, whose resolved binaries depend on the population of Galactic systems, but whose unresolved background is not important for the set of resolved binaries in the Milky Way. Indeed, for such a case in isolation, the formalism here yields the result found in \citet{callister_shouts_2020}. Conversely, this multi-population formalism also naturally accounts for the inverse scenario that may be relevant for the population of extragalactic WD+WD systems, which may give rise to a substantial astrophysical foreground \citep{staelens_likelihood_2024} --- and therefore impact our ability to resolve Galactic sources --- but is not likely to produce individually-resolvable sources. Finally, this result can be generalized to the transdimensional case simply by allowing $P$ to vary, thereby allowing potential unconfirmed subpopulations (for instance, primordial black holes) to be proposed and model comparison on their presence/absence to be included within the overall population model.

\end{document}